\documentclass[a4paper,11pt]{article}
\pdfoutput=1
\usepackage{jcappub}
\usepackage[T1]{fontenc}
\usepackage{subfigure}
\usepackage{hhline}
\usepackage{xcolor}
\usepackage{aas_macros}

\usepackage{amsmath, amssymb}
\usepackage{graphicx}
\usepackage{hyperref}
\usepackage{geometry}
\usepackage{physics}
\usepackage{comment}
\usepackage[T1]{fontenc} 
\usepackage{booktabs}

\newcommand{\be}{\begin{equation}}
\newcommand{\ee}{\end{equation}}
\newcommand{\bea}{\begin{eqnarray}}
\newcommand{\eea}{\end{eqnarray}}
\newcommand{\beq}{\begin{eqnarray}}
\newcommand{\eeq}{\end{eqnarray}}

\title{\boldmath \huge From Fluctuation to Polarization: Imprints of $\mathcal{O}(1-10)\, \mathrm{Mpc}^{-1}$ Curvature Perturbations in CMB B-modes from Scalar-Induced Gravitational Waves}

\author[a]{Aurora Ireland,}
\author[b]{~Kuver Sinha,}
\author[b]{~Tao Xu}
\affiliation[a]{Stanford Institute for Theoretical Physics, Department of Physics, Stanford University, Stanford, CA 94305}
\affiliation[b]{Homer L. Dodge Department of Physics and Astronomy, University of Oklahoma, Norman, OK 73019, USA}

\emailAdd{anireland@stanford.edu}
\emailAdd{kuver.sinha@ou.edu}
\emailAdd{tao.xu@ou.edu}

\abstract{Probing primordial curvature perturbations on small scales, beyond those accessible using cosmic microwave background (CMB) primary anisotropies and Lyman-$\alpha$ forest data, remains a major open challenge. Current constraints on the scalar power spectrum at these scales are either weak (e.g. $\mathcal{P}_\mathcal{R} \lesssim 10^{-4}$ from CMB spectral distortions) or rely heavily on model-dependent assumptions about small-scale structure. In this work, we propose a novel method to probe the small-scale primordial power spectrum using scalar-induced tensor perturbations, which are inevitably sourced by curvature perturbations at second order in cosmological perturbation theory. While induced tensor modes have traditionally been studied in the context of the stochastic gravitational wave background, we highlight a complementary observable: the distinctive pattern of B-mode polarization they imprint on the CMB. We compute the angular spectrum of these B-modes arising from enhanced scalar perturbations and show that the resulting signal can be competitive with inflationary predictions for values of the tensor-to-scalar ratio targeted in upcoming CMB experiments, most notably CMB-Stage 4. We map the region of the scalar power spectrum to which these future B-mode experiments will be sensitive and compare with existing constraints, finding it to exceed current sensitivities at $k \sim \mathcal{O}(1-10)\, \mathrm{Mpc}^{-1}$. In addition to providing a new CMB-based probe of the small-scale power spectrum, this work also motivates dedicated B-mode searches at higher multipoles ($\ell \gtrsim 100$).}

\begin{document}
\maketitle
\flushbottom

\section{Introduction}

A central goal of observational cosmology is to constrain the statistical properties of the primordial perturbations that seeded structure formation in the Universe. The power spectrum of comoving curvature perturbations, $\mathcal{P}_\mathcal{R}(k)$, provides a key window into the physics of inflation and the early Universe. Observations across a wide range of scales have enabled significant progress in measuring this spectrum, yet important gaps remain. On the largest scales, constraints are provided by the cosmic microwave background (CMB) anisotropies as measured by \emph{Planck} and related experiments~\cite{Planck2018}. These measurements pin down $\mathcal{P}_\mathcal{R}(k)$ at wavenumbers $k \lesssim 0.1~\mathrm{Mpc}^{-1}$ to remarkable precision. At smaller scales, Lyman-$\alpha$ forest measurements~\cite{LymanAlpha} extend these constraints up to $k \sim 1~\mathrm{Mpc}^{-1}$, albeit with larger uncertainties.

However, beyond these scales, direct constraints on the primordial power spectrum rapidly weaken. The absence of linear tracers of the primordial fluctuations at $k \gtrsim 10~\mathrm{Mpc}^{-1}$ creates an observational blind spot. Indirect constraints based on spectral distortions of the CMB, specifically $\mu$-type and $y$-type distortions, offer a powerful method to probe $\mathcal{P}_\mathcal{R}(k)$ at smaller scales~\cite{Chluba:2012gq,Chluba:2012we}. Current bounds from COBE/FIRAS~\cite{COBEFIRAS1,COBEFIRAS2} already limit $\mathcal{P}_\mathcal{R}(k) \lesssim 10^{-4}$ at $k \sim 10^2$ -- $10^4~\mathrm{Mpc}^{-1}$, but are unable to access amplitudes below this level. A proposed mission such as PIXIE~\cite{PIXIE,Kogut:2024} could improve these limits by up to two orders of magnitude. Currently, the prospects of PIXIE remain uncertain, leaving a critical gap in our ability to constrain the primordial power spectrum on small scales. 

Indeed, probing amplitudes of $\mathcal{P}_\mathcal{R}(k) \lesssim 10^{-5}$ across the range $k \sim 10$ -- $10^5~\mathrm{Mpc}^{-1}$ remains extremely challenging. Standard CMB temperature and polarization anisotropy measurements lose sensitivity at smaller scales due to Silk damping and finite resolution. Large-scale structure (LSS) surveys traditionally probe curvature perturbations through the associated density fluctuations. However, observing low-mass halos at these scales is difficult, and the non-linear evolution of structures introduces complex dependencies that limit the ability to reconstruct the primordial power spectrum. Furthermore, there exists an observational bifurcation: if the amplitude of the primordial perturbations is large enough to trigger non-linear structure formation, the resulting compact objects (e.g., primordial black holes) are observable; otherwise, if subcritical, the induced structures are diffuse and escape detection. Thus, while primordial black holes~\cite{PBHrev} can potentially constrain the primordial power spectrum, forming these objects requires an extreme enhancement of $\mathcal{P}_\mathcal{R}(k)$, typically to levels of order $10^{-2}$. The corresponding constraints are thus limited to regions of parameter space where the primordial power spectrum is already dramatically amplified. By contrast, the ability to probe smaller amplitude fluctuations remains an open frontier. 

Constraints on $\mathcal{P}_{\mathcal{R}}(k)$ at small scales have recently been derived from the gravitational heating of stars in ultra-faint dwarf galaxies, which limits the abundance of dark matter substructure~\cite{Graham2024}. While this method offers strong limits in the range $k\sim 10$ -- $10^3\,\mathrm{Mpc}^{-1}$, comparable to future spectral distortion sensitivity from PIXIE, it depends on uncertain modeling of sub-halo survival, tidal disruption, and galaxy selection biases. Moreover, translating primordial fluctuations to present-day substructure populations requires assumptions about non-linear collapse and tidal evolution, which depend on the underlying cosmology, dark matter physics, and the galactic environment. $\mathcal{P}_{\mathcal{R}}(k)$ can also be constrained on the basis that fluctuations large enough to trigger early structure formation impact the CMB and other observables. This direction was recently pursued by the authors of \cite{Qin:2025ymc}, who derived bounds on $\mathcal{P}_{\mathcal{R}}(k)$ based on early reionization, and by \cite{Bringmann:2025cht}, who  constrained the formation of ultra-compact mini-halos based on microlensing and heating of
the CMB due to baryonic accretion.
These methods provide valuable but model-dependent probes of the primordial power spectrum.

To overcome these challenges, novel methods must be employed. One promising strategy involves tracking how primordial fluctuations affect particle distributions. Nonetheless, the radiation-dominated era presents obstacles: pressure support in the Standard Model plasma erases small-scale inhomogeneities once they enter the horizon. Additionally, surviving perturbations predominantly retain their scalar character, making them difficult to disentangle from late-time astrophysical effects.  This motivates the search for a signal that is preserved through plasma interactions and is statistically distinct from scalar modes. 

In this work, we propose a complementary approach based on induced gravitational waves (GWs). Scalar perturbations inevitably generate tensor modes at second order in cosmological perturbation theory~\cite{Ananda2007,Baumann2007}. These GWs retain information about the primordial curvature spectrum without being erased by the hot plasma, and their tensorial nature allows them to be distinguished observationally. However, detecting induced GWs at $k \sim \mathcal{O}(1-10)\, \mathrm{Mpc}^{-1}$ itself poses significant challenges. Direct detection through GW observatories targets much higher frequency ranges, corresponding to smaller scales. Pulsar timing arrays (PTAs) \cite{NANOGrav:2023gor, EPTA:2023fyk, Reardon:2023gzh, Xu:2023wog} and future detectors like LISA \cite{LISA:2017pwj} are sensitive to $k \sim 10^6 \,\mathrm{Mpc}^{-1}$ and $10^{12}\,\mathrm{Mpc}^{-1}$ respectively, orders of magnitude larger than the scales of interest here. Contributions of induced GWs to the effective number of relativistic species ($N_\mathrm{eff}$) offer an alternative indirect probe, but at the cost of losing spectral information and facing degeneracies with other relativistic relics such as light dark matter candidates.

While the direct detection of induced GWs in this frequency band is beyond our current capabilities, fortunately this is not the only observable where scalar-induced tensor modes can manifest. Indeed, these same tensor perturbations also imprint a characteristic B-mode polarization pattern in the CMB. Historically, detecting a B-mode signal (above astrophysical foregrounds) been regarded as a smoking gun for GWs produced during inflation. More recently, however, Ref.~\cite{Greene:2024} has robustly demonstrated that contrary to popular lore, tensor perturbations from non-inflationary sources in the early universe can actually produce B-mode amplitudes rivaling those from inflation on observable scales. The authors of Ref.~\cite{Greene:2024} considered tensor modes from a first-order cosmological phase transition occurring in the ``late'' but pre-recombination universe as a proof of principle that one can have sufficient power in the tensor power spectrum to affect the scales probed by the CMB. 

Here, we consider scalar-induced tensor modes and demonstrate that these too can be used to probe nominally smaller scales than CMB primary anisotropies. The B-modes generated by induced GWs present an opportunity to access primordial curvature perturbations at intermediate scales, complementary to traditional probes. Moreover, they offer a clean probe which is statistically distinct compared to scalar perturbations, which at linear order produce only E-mode polarization. Consequently, B-mode polarization provides a relatively uncontaminated window into the tensor sector, particularly when combined with delensing techniques to reduce contamination from lensing-induced B-modes. 

Our main result is that induced GWs, arising inevitably from scalar fluctuations, source B-mode polarization in the CMB which provides an indirect probe of the scalar power spectrum $\mathcal{P}_\mathcal{R}(k)$ on scales beyond those accessible via Lyman-$\alpha$ forest or CMB spectral distortion measurements (see Fig.~\ref{fig:Pkconstraints}). Importantly, this method leverages future CMB polarization surveys such as the CMB-Stage 4 (CMB-S4) experiment~\cite{CMBS4,CMBS4:2020} and also provides additional motivation beyond de-lensing for dedicated B-mode searches in the higher ($\ell \gtrsim 100$) multipole range. 

Our strategy will be to first compute the induced tensor power spectrum $\mathcal{P}_h(k)$ corresponding to a given form for the primordial scalar power spectrum $\mathcal{P}_\mathcal{R}(k)$. Next, we derive the contribution of $\mathcal{P}_h(k)$ to temperature anisotropies and the CMB polarization tensor, from which one can extract the angular spectrum of B-mode polarization $\mathcal{D}_\ell^{BB}$. Note that our implementation leverages a combination of analytic results and numerics, using the Boltzmann solver \texttt{CLASS}~\cite{CLASS:I,CLASS:II}, in order to maximize precision.

Using this machinery, we compute $\mathcal{D}_\ell^{BB}$ for induced tensor modes presuming narrowly peaked scalar spectra and compare with inflationary predictions for target values of the tensor-to-scalar ratio $r$. For peak values in the range $k_p \sim \mathcal{O}(1-10) \, \text{Mpc}^{-1}$, we find the signals to be competitive\footnote{More precisely, they yield the same signal-to-noise ratios.} with inflationary models with $r = 10^{-3}$. We map these contours to the $\mathcal{P}_\mathcal{R}(k)$ plane in order to highlight regions of future sensitivity from current and upcoming CMB polarization surveys. A detection or non-observation of scalar induced B-modes in these future experiments will allow us to infer information about $\mathcal{P}_\mathcal{R}$, complementing existing and proposed spectral distortion constraints.

This work is organized as follows: We begin in Sec.~\ref{sec:theory} by reviewing the generation of tensor perturbations from scalar fluctuations at second order in cosmological perturbation theory, including deriving $\mathcal{P}_h(k)$ for generic forms of the scalar power spectrum $\mathcal{P}_\mathcal{R}(k)$. In Sec.~\ref{sec:Bmode}, we review the formalism for computing the angular spectrum of B-mode polarization $\mathcal{D}_\ell^{BB}$ giving an initial tensor power spectrum $\mathcal{P}_h(k)$. We evaluate this explicitly for a sharply peaked $\mathcal{P}_\mathcal{R}$ and compare the resulting spectra with those predicted by inflation with a given $r$. In Sec.~\ref{sec:prospects}, we discuss observational prospects for either detecting or constraining the signal with current and future CMB experiments. We comment also on the induced GW signal as a complementary probe. We conclude in Sec.~\ref{sec:conclusions} with a summary of our results and suggestions for future directions.

\section{Scalar-Induced Tensor Perturbations}
\label{sec:theory}

\subsection{Theory}

In the context of cosmological perturbation theory, scalar (curvature) perturbations can source tensor perturbations (GWs) at second order. This phenomenon leads to the generation of a stochastic background of gravitational waves, even in the absence of primordial tensor modes from inflation. The study of these induced gravitational waves provides a unique window into the small-scale behavior of the primordial curvature perturbations $\mathcal{R}$ beyond the scales probed by CMB temperature anisotropies.

The evolution of tensor perturbations $h_{ij}$ in a flat Friedmann–Lemaître–Robertson–Walker (FLRW) universe is governed by the perturbed Einstein equations. At second order in cosmological perturbation theory, the equation for the Fourier mode $h_{\mathbf{k}}^{\lambda}(\eta)$ with polarization $\lambda$ is given by
\begin{equation}\label{eq:tensor_eom}
h''_{\mathbf{k},\lambda}(\eta) + 2 \mathcal{H} h'_{\mathbf{k},\lambda}(\eta) + k^2 h_{\mathbf{k},\lambda}(\eta) = S_{\mathbf{k}}^{\lambda}(\eta) \,,
\end{equation}
where primes denote derivatives with respect to conformal time $\eta$, $\mathcal{H} = a'/a$ is the conformal Hubble parameter, and $S_{\mathbf{k}}^{\lambda}(\eta)$ is the source term arising from products of first-order scalar perturbations. Explicitly, it takes the form
\begin{equation}\label{eq:source_term}
    S_{\mathbf{k}}^{\lambda}(\eta) = \int \frac{d^3 q}{(2\pi)^{3/2}} Q_\lambda(\mathbf{k}, \mathbf{q}) f(\eta, |\mathbf{k} - \mathbf{q}|, q) \mathcal{R}(\mathbf{k} - \mathbf{q}) \mathcal{R}(\mathbf{q}) \,,
\end{equation}
where $Q_\lambda(\mathbf{k}, \mathbf{q}) = \varepsilon_{\lambda}^{ij}(\hat{\mathbf{k}}) q_i q_j$ is a projection factor, $\mathcal{R}(\mathbf{k})$ is the primordial curvature perturbation, and $f(\eta, |\mathbf{k} - \mathbf{q}|, q)$ encodes the time evolution and mode couplings between these scalar perturbations. Implicit in deriving this expression, we have assumed vanishing anisotropic stress $\Psi = \Phi$ and decomposed the Newtonian gauge gravitational potential $\Phi(\eta,\mathbf{k})$ into a primordial curvature perturbation $\mathcal{R}(\mathbf{k})$ and the transfer function $\phi(k \eta)$,
\begin{equation}
    \Phi(\eta,\mathbf{k}) = \frac{3 + 3w}{5 + 3w} \phi(k \eta) \mathcal{R}(\mathbf{k}) \,,
\end{equation}
with $w$ the equation of state. This decomposition is convenient because it separates out the effects of the initial statistical correlations, as described by the scalar power spectrum,\footnote{We denote \textit{dimensionful} power spectra as $P(k)$ and \textit{dimensionless} power spectra as $\mathcal{P}(k) = \frac{k^3}{2\pi^2} P(k)$.}
\begin{equation}\label{eq:PRdef}
    \langle \mathcal{R}(\mathbf{k}) \mathcal{R}(\mathbf{k}') \rangle = P_\mathcal{R}(k) \delta^{(3)}(\mathbf{k} + \mathbf{k}') \,,
\end{equation}
from the deterministic evolution encoded by the scalar perturbation transfer function $\phi(k \eta)$. In the absence of entropy perturbations, this latter object evolves according to the equation
\begin{equation}\label{eq:TransferEq}
    \phi''(k \eta) + \frac{6(1+w)}{(1+3w)} \frac{1}{\eta} \phi'(k \eta) + w k^2 \phi(k \eta) = 0 \,,
\end{equation}
where again primes denote derivatives with respect to conformal time $\eta$. The transfer function enters in the definition of the function $f(\eta, |\mathbf{k} - \mathbf{q}|, q)$ appearing in Eq.~(\ref{eq:source_term}), 
\begin{equation}\label{eq:fdef}
\begin{split}
    f(\eta, p, q) = \frac{3(1+w)}{(5+3w)^2} \bigg[ 2(5+ & 3w) \phi(p \eta) \phi(q \eta) + \eta^2 (1+3w)^2 \phi'(p \eta) \phi'(q \eta)\\
    & + 2 \eta(1+3w) \big(\phi(p \eta) \phi'(q\eta) + \phi'(p \eta) \phi(q \eta) \big) \bigg] \,,
\end{split}
\end{equation}
where $p = |\mathbf{k} - \mathbf{q}|$ and we have used also that $\mathcal{H} = \frac{2}{(1 + 3w)} \frac{1}{\eta}$.

Generically, one can solve Eq.~(\ref{eq:tensor_eom}) via the method of Green's functions to obtain
\begin{equation}\label{eq:hsol}
    h_{\mathbf{k}}^{\lambda}(\eta) = \frac{4}{a(\eta)} \int_{\eta_i}^{\eta} d\bar{\eta} \, G_{\mathbf{k}}(\eta,\bar{\eta}) a(\bar{\eta}) S_{\mathbf{k}}^{\lambda}(\bar{\eta}) \,,
\end{equation}
where $G_{\mathbf{k}}(\eta,\bar{\eta})$ is a Green's function for the homogeneous equation with delta function source, 
\begin{equation}\label{eq:GreenEq}
    G''_{\mathbf{k}}(\eta,\bar{\eta}) + \left( k^2 - \frac{a''(\eta)}{a(\eta)} \right) G_{\mathbf{k}}(\eta,\bar{\eta}) = \delta(\eta - \bar{\eta}) \,.
\end{equation}
Substituting the explicit form for the source $S_{\mathbf{k}}^\lambda$ of Eq.~(\ref{eq:source_term}) into Eq.~(\ref{eq:hsol}) and isolating the time integral, it is convenient to define the combination
\begin{equation}\label{eq:Idef}
    \mathcal{I}(\eta, |\mathbf{k} - \mathbf{q}|,q) = \frac{1}{a(\eta)} \int_{0}^\eta d\bar{\eta} \, G_{\mathbf{k}}(\eta,\bar{\eta}) a(\bar{\eta}) f(\bar{\eta}, |\mathbf{k} - \mathbf{q}|,q) \,.
\end{equation}
This kernel then appears in the solution for $h_{\mathbf{k}}^\lambda$ as
\begin{equation}\label{eq:exactsol}
    h_{\mathbf{k}}^\lambda(\eta) = 4 \int \frac{d^3 q}{(2\pi)^{3/2}} Q_\lambda(\mathbf{k}, \mathbf{q}) \mathcal{I}(\eta, |\mathbf{k} - \mathbf{q}|, q) \mathcal{R}(\mathbf{k} - \mathbf{q}) \mathcal{R}(\mathbf{q}) \,.
\end{equation}
The equal-time 2-point correlation function of $h_{\mathbf{k}}^\lambda$ reads
\begin{equation}
\begin{split}
    \langle h_{\mathbf{k}}^\lambda(\eta) h_{\mathbf{k}'}^{\lambda'}(\eta) \rangle & = 16 \! \int \!\! \frac{d^3 q_1}{(2\pi)^{3/2}} \! \int \!\! \frac{d^3 q_2}{(2\pi)^{3/2}} Q_\lambda(\mathbf{k}, \mathbf{q}_1) Q_{\lambda'}( \mathbf{k}', \mathbf{q}_2) \\
    & \times \mathcal{I}(\eta, |\mathbf{k} - \mathbf{q}_1|, q_1) \mathcal{I}(\eta, |\mathbf{k}' - \mathbf{q}_2|, q_2) \langle \mathcal{R}(\mathbf{k} - \mathbf{q}_1) \mathcal{R}(\mathbf{q}_1) \mathcal{R}(\mathbf{k}' - \mathbf{q}_2) \mathcal{R}(\mathbf{q}_2) \rangle \,, 
\end{split}
\end{equation}
which features a 4-point function of the primordial curvature perturbation $\mathcal{R}$. This 4-point correlator can generically be decomposed into a connected and disconnected contributions, where the connected component is defined in terms of the trispectrum $\mathcal{T}_{\mathcal{R}}(\mathbf{k}_1, \mathbf{k}_2, \mathbf{k}_3, \mathbf{k}_4)$ and the disconnected component can be decomposed using Wick's theorem. We will focus on the case of Gaussian perturbations, in which case the connected trispectrum vanishes we need only worry about the disconnected contribution. There are three contractions in the decomposition, of which one is vanishing and the other two can be shown to give the same contribution (upon using the properties of the delta functions and projection factors to simplify). Using one of the delta functions to perform the integral over $\mathbf{q}_2$ and renaming $\mathbf{q}_1 \rightarrow \mathbf{q}$, the result is
\begin{equation}
    \langle h_{\mathbf{k}}^\lambda(\eta) h_{\mathbf{k}'}^{\lambda'}(\eta) \rangle \! = \! 32 \! \int \!\! \frac{d^3 q}{(2\pi)^{3}} Q_\lambda(\mathbf{k}, \mathbf{q})^2 \mathcal{I}(\eta, |\mathbf{k} - \mathbf{q}|, q)^2 P_\mathcal{R}(|\mathbf{k} - \mathbf{q}|) P_\mathcal{R}(q) \delta_{\lambda \lambda'} \delta^{(3)}(\mathbf{k} + \mathbf{k}') \,,
\end{equation}
with the scalar power spectrum $P_\mathcal{R}$ defined in Eq.~(\ref{eq:PRdef}). Upon comparing this with the definition of the (dimensionless) tensor power spectrum
\begin{equation}\label{eq:Phdef}
    \langle h_{\mathbf{k}}^\lambda h_{\mathbf{k}'}^{\lambda'} \rangle = \frac{2 \pi^2}{k^3} \mathcal{P}_h^\lambda(\eta, k) \delta_{\lambda \lambda'} \delta^{(3)}(\mathbf{k} + \mathbf{k}') \,,
\end{equation}
we may identify 
\begin{equation}\label{eq:Phsol}
    \mathcal{P}_h^\lambda(\eta, k) = \left( \frac{k^3}{2 \pi^2} \right) 32 \! \int \!\! \frac{d^3 q}{(2\pi)^{3}} Q_\lambda(\mathbf{k}, \mathbf{q})^2 \mathcal{I}(\eta, |\mathbf{k} - \mathbf{q}|, q)^2 P_\mathcal{R}(|\mathbf{k} - \mathbf{q}|) P_\mathcal{R}(q) \,.
\end{equation}
Explicitly evaluating this expression for our case of interest will be the goal of the next two sub-sections.

\subsection{Explicit Evaluation}

We will focus on scalar-induced tensor perturbations produced in the radiation dominated era, during which $w = 1/3$, $a = \eta$, and $\mathcal{H} = 1/\eta$. Solving Eq.~(\ref{eq:GreenEq}) during radiation domination with initial conditions\footnote{Without loss of generality we set $\eta_i = 0$.} $h_{\mathbf{k},\lambda}(0) = h_{\mathbf{k},\lambda}'(0) = 0$, we obtain the Green's function 
\begin{equation}\label{eq:Green}
    G_{\mathbf{k}}^{\rm RD}(\eta,\bar{\eta}) = \frac{\sin[k(\eta - \bar{\eta})]}{k} \,.
\end{equation}
Meanwhile, solving Eq.~(\ref{eq:TransferEq}) during this era with superhorizon initial conditions $\phi(0) = 1$, $\phi'(0) = 0$ yields
\begin{equation}
    \phi_{\rm RD}(k \eta) = \frac{3 \sqrt{3}}{k \eta} j_1\left( \frac{k \eta}{\sqrt{3}} \right) \,,
\end{equation}
where $j_1(x) = \frac{\sin(x)}{x^2} - \frac{\cos(x)}{x}$ is a spherical Bessel function of the first kind. Substituting this solution into Eq.~(\ref{eq:fdef}) with $w = 1/3$ then gives an explicit expression for $f(\eta, |\mathbf{k} - \mathbf{q}|,q)$,
\begin{equation}\label{eq:fRD}
\begin{split}
    & f_{\rm RD}(\eta, p, q) = \frac{12}{p^3 q^3 \eta^6} \bigg[ 18 p q \eta^2 \cos \! \left(\! \frac{p \eta}{\sqrt{3}} \! \right) \cos\! \left(\! \frac{q \eta}{\sqrt{3}} \!\right) + 2 \sqrt{3} p \eta (q^2 \eta^2 - 9) \cos \!\left( \!\frac{p \eta}{\sqrt{3}} \!\right) \sin\! \left( \!\frac{q \eta}{\sqrt{3}} \!\right) \\
    & + 2 \sqrt{3} q \eta (p^2 \eta^2 - 9) \cos\! \left(\! \frac{q \eta}{\sqrt{3}}\! \right) \sin\! \left( \!\frac{p \eta}{\sqrt{3}} \!\right) + (54 - 6(p^2 + q^2) \eta^2 + p^2 q^2 \eta^4) \sin\! \left( \!\frac{p \eta}{\sqrt{3}} \!\right) \sin\! \left( \!\frac{q \eta}{\sqrt{3}} \!\right) \bigg] \,,
\end{split}
\end{equation}
where again for convenience we have written $p = |\mathbf{k} - \mathbf{q}|$.

Upon specifying a form for the scalar power spectrum, this is in principle enough information to compute $\mathcal{I}(\eta, |\mathbf{k} - \mathbf{q}|,q)$ in Eq.~(\ref{eq:Idef}) and finally $\mathcal{P}_h^\lambda(k)$ in Eq.~(\ref{eq:Phsol}). To perform the integral over $\mathbf{q}$, we adopt a coordinate system with $\mathbf{k}$ aligned along the $\hat{z}$ axis, $\mathbf{k} = k(0,0,1)$, and $\mathbf{q}$ defined relative to it, $\mathbf{q} = q(\sin\theta \cos \phi, \sin \theta \sin \phi, \cos \theta)$. The projection factor $Q_\lambda(\mathbf{k}, \mathbf{q}) = \varepsilon_{\lambda}^{ij}(\hat{\mathbf{k}}) q_i q_j$ is then
\begin{equation}
    Q_\lambda(\mathbf{k}, \mathbf{q}) = \frac{q^2}{\sqrt{2}} \sin^2 \theta \times \begin{cases} \cos 2 \phi & \lambda = + \,, \\ \sin 2 \phi & \lambda = \times \,. \end{cases}
\end{equation}
Summing over both polarizations, the total tensor power spectrum $\mathcal{P}_h(k) \equiv \sum_{\lambda = +,\times} \mathcal{P}_h^\lambda(k)$ written in terms of these variables becomes
\begin{equation}\label{eq:Phtot}
    \mathcal{P}_h(\eta, k) = \left( \frac{k^3}{2 \pi^2} \right) \frac{4}{\pi^2} \! \int_0^\infty \!\! dq \, q^6 \int_{-1}^1 \!\! d(\cos \theta) \sin^4 \, \theta \, \mathcal{I}(\eta, |\mathbf{k} - \mathbf{q}|, q)^2 \, P_\mathcal{R}(|\mathbf{k} - \mathbf{q}|) P_\mathcal{R}(q) \,.
\end{equation}
where $|\mathbf{k} - \mathbf{q}| = \sqrt{k^2 + q^2 - 2 k q \cos \theta}$. 

In practice, however, the integrals as written are often numerically intractable. This motivates us to introduce the change of variables
\begin{equation}
    u \equiv \frac{|\mathbf{k} - \mathbf{q}|}{k} \,, \,\,\, v \equiv \frac{q}{k} \,.
\end{equation}
By writing $u = \sqrt{1 + v^2 - 2 v \cos \theta}$, we see that it takes values $u \in [|1-v|, 1+v]$ while $v \in [0, \infty)$. Using also that the Jacobian for this transformation is $J = - k u/v$, the integration measure becomes
\begin{equation}\label{eq:measuretransform}
    \int d^3 q = k^3 \int_0^\infty dv \, v \int_{|1-v|}^{1+v} du \, u \int_0^{2\pi} d\phi \,.
\end{equation}
In terms of $u$ and $v$, the sum over (squared) projection factors is
\begin{equation}\label{eq:Qtransform}
    \sum_{\lambda = +, \times} Q_\lambda(\mathbf{k}, \mathbf{q})^2 = \frac{k^4}{32} \bigg( 4 v^2 - (1 + v^2 - u^2)^2 \bigg)^2 \,.
\end{equation}
We can also express the product of (dimensionful) tensor power spectra in terms of the (dimensionless) tensor power spectra in these new variables,
\begin{equation}\label{eq:PPtransform}
    P_\mathcal{R}(|\mathbf{k} - \mathbf{q}|) P_\mathcal{R}(q) = \frac{4 \pi^4}{k^6 u^3 v^3} \mathcal{P}_\mathcal{R}(u) \mathcal{P}_\mathcal{R}(v) \,.
\end{equation}
Finally, we introduce the dimensionless combination $x \equiv k \eta$ and define the rescaled $\tilde{\mathcal{I}} \equiv k^2 \mathcal{I}$, with $\mathcal{I}$ in Eq.~(\ref{eq:Idef}). In terms of the new variables, this rescaled kernel during radiation domination is\footnote{Our $\tilde{\mathcal{I}}_{\rm RD}$ coincides with the function $I_{\rm RD}$ of Ref.~\cite{Kohri:2018}.}
\begin{equation}
    \tilde{\mathcal{I}}_{\rm RD}(x, u, v) = \frac{1}{x} \int_0^x d \bar{x} \, \sin(x - \bar{x}) \, \bar{x} \, f_{\rm RD}(\bar{x}, u, v) \,,
\end{equation}
with $f_{\rm RD}$ in Eq.~(\ref{eq:fRD}). It is actually possible to evaluate the integral over $\bar{x}$ appearing in $\mathcal{I}_{\rm RD}$ analytically, as was first demonstrated in Refs.~\cite{Kohri:2018,Espinosa:2018}. We have verified that we recover the same expression, which we reproduce below
\begin{equation}\label{eq:analyticI}
\begin{split}
    \tilde{\mathcal{I}}_{\rm RD}(x, u, v) = & \frac{9}{u^3 v^3 x^4} \bigg[ 2 u v x^2 \cos \! \left( \! \frac{u x}{\sqrt{3}} \! \right) \cos \! \left( \! \frac{v x}{\sqrt{3}} \! \right) - 2 \sqrt{3} x u \cos \! \left( \! \frac{u x}{\sqrt{3}} \! \right) \sin \! \left( \! \frac{u x}{\sqrt{3}} \! \right) \\
    & - 2 \sqrt{3} x v \sin \! \left( \! \frac{u x}{\sqrt{3}} \! \right) \cos \! \left( \! \frac{v x}{\sqrt{3}} \! \right) + (6 + (u^2 + v^2 - 3) x^2) \sin \! \left( \! \frac{u x}{\sqrt{3}} \! \right) \sin \! \left( \! \frac{v x}{\sqrt{3}} \! \right) \bigg] \\
    & + \frac{3(u^2 + v^2 - 3)}{4 u^3 v^3 x} \bigg[ \sin x \, F_{Ci}(x,u,v) + \cos x \, F_{Si}(x, u, v) \bigg] \,,
\end{split}
\end{equation}
where we have defined the functions
\begin{equation}
\begin{split}
    F_{Ci}(x,u,v) = & \text{Ci} \left( \! \left( \! 1 + \frac{u - v}{\sqrt{3}} \right) \! x \! \right) + \text{Ci} \left( \! \left( \! 1 - \frac{u - v}{\sqrt{3}} \right) \! x \! \right) \\ 
    & - \text{Ci} \left( \! \left( \! 1 + \frac{u + v}{\sqrt{3}} \right) \! x \! \right) - \text{Ci} \left( \bigg| 1 - \frac{u + v}{\sqrt{3}} \bigg| x \! \right) + \ln \bigg| \frac{3 - (u+v)^2}{3 - (u - v)^2} \bigg| - 4 u v \,,
\end{split}
\end{equation}
and
\begin{equation}
\begin{split}
    F_{Si}(x,u,v) = & \text{Si} \left( \! \left( \! 1 + \frac{u + v}{\sqrt{3}} \right) \! x \! \right) + \text{Si} \left( \! \left( \! 1 - \frac{u + v}{\sqrt{3}} \right) \! x \! \right) \\ 
    & - \text{Si} \left( \! \left( \! 1 + \frac{u - v}{\sqrt{3}} \right) \! x \! \right) - \text{Si} \left( \! \left( \! 1 - \frac{u - v}{\sqrt{3}} \right) \! x \! \right) \,,
\end{split}
\end{equation}
with $\text{Ci}(z)$ and $\text{Si}(z)$ cosine and sine integrals respectively,
\begin{equation}
    \text{Ci}(z) = - \int_z^\infty \!\! d \bar{z} \, \frac{\cos \bar{z}}{\bar{z}} \,, \,\,\, \text{Si}(z) = \int_0^z \!\! d\bar{z} \, \frac{\sin \bar{z}}{\bar{z}} \,.
\end{equation}
Combining Eqs.~(\ref{eq:measuretransform}), (\ref{eq:Qtransform}), and (\ref{eq:PPtransform}) in Eq.~(\ref{eq:Phsol}) and summing over polarizations, we finally obtain the total tensor power spectrum in terms of these new variables,\footnote{In the literature, one typically performs a second change of variables, introducing $s \equiv u - v$ and $t \equiv u + v - 1$. In these variables, the integration domain becomes rectangular, simplifying the numerical implementation. For the scalar power spectra considered here, however, working in terms of $u$ and $v$ is sufficient.}
\begin{equation}\label{eq:Phuv}
    \mathcal{P}_h(x, k) = \frac{1}{2} \int_0^\infty \!\! dv \, \int_{|1-v|}^{1+v} \!\! du \left( \frac{4v^2 - (1 + v^2 - u^2)^2}{u v} \right)^2 \, \tilde{\mathcal{I}}_{\rm RD}(x,u,v)^2 \, \mathcal{P}_\mathcal{R}(u) \mathcal{P}_\mathcal{R}(v) \,,
\end{equation}
with $\tilde{\mathcal{I}}_{\rm RD}(x,u,v)$ given by Eq.~(\ref{eq:analyticI}) above.

After production, tensor perturbations redshift as radiation while their wavelengths remain inside the horizon. During the radiation-dominated epoch, their energy density is given by
\begin{equation}\label{eq:OmegaGW}
    \Omega_{\rm GW}(x, k)=\frac{x^2}{24} \, \overline{\mathcal{P}_h(x, k)} \,.
\end{equation}
Here $\overline{\mathcal{P}_h(x, k)}$ denotes the time-averaged tensor power spectrum, accounting for the oscillations of $\tilde{\mathcal{I}}_{\rm RD}(x,u,v)$.
The present-day GW energy density is thus
\begin{equation}\label{eq:OmegaGWtoday}
    \Omega_{\rm GW}^0(k) = \Omega_{\text{r},0} \, \Omega_{\rm GW}(x\gg 1, k) \,.
\end{equation}
where $\Omega_{\rm GW}(x\gg 1, k)$ is evaluated when the tensor mode is well inside the horizon and the source term has decayed, and $\Omega_{\text{r},0}$ denotes the energy density parameter of radiation at present.

\subsection{Sample Power Spectra}

We now proceed to evaluate Eq.~(\ref{eq:Phuv}) for simple forms of the scalar power spectrum. We begin with the case of a Dirac delta function peaked at some particular scale $k_p$, $\mathcal{P}_\mathcal{R}(k) = \mathcal{A} \delta(\ln(k/k_p))$, for which the integrals over $u$ and $v$ in Eq.~(\ref{eq:Phuv}) can be evaluated exactly. The existence of an analytic form for the induced tensor spectrum $\mathcal{P}_h(k)$ makes this case especially instructive for understanding how a narrow feature in the scalar power spectrum might contribute to induced tensor perturbations. Examining the time evolution of $\mathcal{P}_h(k)$ in this idealized scenario will also provide insight into the appropriate time to evaluate the ``initial'' tensor power spectrum, relevant for our later calculations of the B-mode amplitude. 

The delta function has one unsatisfactory feature, however, which is that the infrared (IR) tail of the tensor power spectrum scales as $\mathcal{P}_h \propto k^2$, rather than the expected causality-limited scaling $\mathcal{P}_h^{\rm IR} \propto k^3$. This is of course unphysical, and so next we consider the case of a ``box function'' power spectrum, which remains semi-analytically tractable while ensuring physically consistent behavior in the IR.  

\subsubsection{Delta Function}

We can approximate a sufficiently narrow scalar power spectrum by a delta function\footnote{Keep in mind that $\mathcal{P}_\mathcal{R}$ is dimensionless, and so the argument of the delta function should be chosen to be dimensionless.} peaked at one particular comoving wavenumber $k_p$,
\begin{equation}
    \mathcal{P}_\mathcal{R}^\delta (k) = \mathcal{A} \delta( \ln(k/k_p) ) = \mathcal{A} k \delta(k - k_p) \,,
\end{equation}
where $\mathcal{A}$ the amplitude of the peak and we have used the property $\delta(f(x)) = \sum_i \frac{\delta(x - x_i)}{|f'(x_i)|}$ in writing the second equality. In terms of the variables $u = |\mathbf{k} - \mathbf{q}|/k$ and $v = q/k$, the product $\mathcal{P}_\mathcal{R}^\delta(u) \mathcal{P}_\mathcal{R}^\delta(v)$ appearing in Eq.~(\ref{eq:Phuv}) is
\begin{equation}
    \mathcal{P}_\mathcal{R}^\delta(u) \mathcal{P}_\mathcal{R}^\delta(v) = \mathcal{A}^2 u v \, \delta\left(u - \frac{k_p}{k} \right) \delta\left(v - \frac{k_p}{k} \right) \,,
\end{equation}
in the delta function case. Substituting this into Eq.~(\ref{eq:Phuv}) and using the delta functions to carry out the integrals over $u$ and $v$, we obtain
\begin{equation}\label{eq:Phdelta}
    \mathcal{P}_h^\delta(x,k) = \mathcal{A}^2 \frac{(4 k_p^2 - k^2)^2}{2 k^2 k_p^2} \, \tilde{\mathcal{I}}_{\rm RD}\!\left( x, \frac{k_p}{k}, \frac{k_p}{k} \right)^2 \Theta(2 k_p - k) \,,
\end{equation}
where $\Theta(2 k_p - k)$ is the Heaviside step function enforcing $k \leq 2 k_p$. This condition on $k$ comes from performing the integral over $u$; since the delta functions set $u = v = k_p/k$ while the integration limits restrict $u = k_p/k \geq |1 - v| = |1 - k_p/k|$, we are forced to take $k \leq 2 k_p$. Physically, the boundaries in the integral over $u$ correspond to the conservation of momentum, and so demanding $k \leq 2 k_p$ is equivalent to respecting momentum conservation. 

\begin{figure}[t!]
    \centering
    \includegraphics[width=1.0\textwidth]{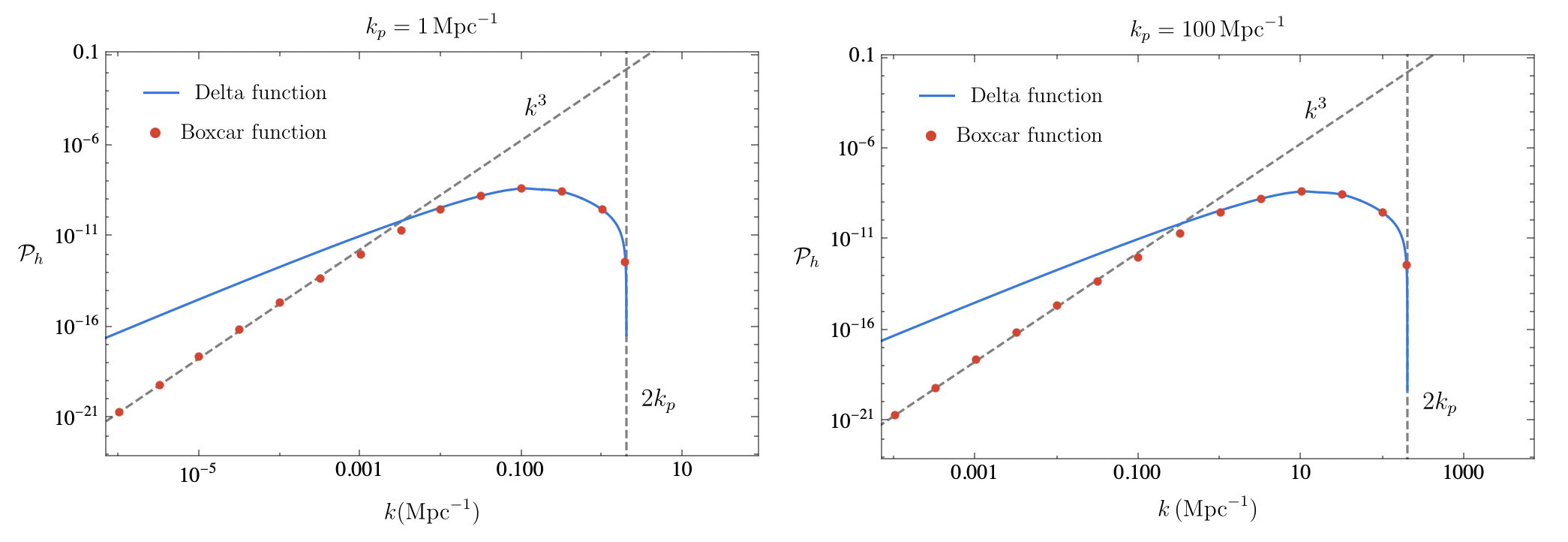}\qquad
    \caption{Sample tensor power spectra $\mathcal{P}_h(k)$ evaluated at horizon crossing ($x=1$) for tensor perturbations induced by delta function (blue) and boxcar function (red) peaks in the scalar power spectrum $\mathcal{P}_\mathcal{R}(k)$. We consider peaks at $k_p = 1 \, \text{Mpc}^{-1}$ (left) and $k_p = 100 \, \text{Mpc}^{-1}$ (right) and fix the amplitude as $\mathcal{A} = 10^{-4.4}$, corresponding to the limit set by CMB spectral distortions in this range of wavenumbers. For the boxcar function, we take the width to be quite narrow, $\Delta = 0.01$. The gray dashed vertical line indicates the cutoff at $k = 2 k_p$ corresponding to momentum conservation. The gray dashed line labeled $k^3$ shows the expected IR scaling. Note that the delta function spectrum falls off as $k^2$ in the IR, which is unphysical, whereas the boxcar function spectrum demonstrates the correct $k^3$ scaling expected of a causal source.}
    \label{fig:tensorPk}
\end{figure}

Thus we see that at large $k > 2 k_p$ the spectrum is required to vanish. In the opposite regime of small $k \ll k_p$, we can expand the function $\tilde{\mathcal{I}}_{\rm RD}(x,k_p/k, k_p/k)$ to obtain
\begin{equation}
    \lim_{k/k_p \ll 1} \tilde{\mathcal{I}}_{\rm RD}(x,k_p/k, k_p/k) = c_2 \frac{k^2}{k_p^2} + c_4 \frac{k^4}{k_p^4} + c_5 \frac{k^5}{k_p^5} + c_6 \frac{k^6}{k_p^6} \,.
\end{equation}
The explicit forms of the coefficients $c_{n}(x,k/k_p)$, with $n=2,4,5,6$, are not very important since $k/k_p$ only appears in the argument of either sine or cosine functions or logarithms; the leading scaling behavior comes from the $k^n$ prefactors which we have extracted. In particular, the leading-order behavior for small $k \ll k_p$ is $\tilde{\mathcal{I}}_{\rm RD} \sim k^2$. The entire expression in Eq.~(\ref{eq:Phdelta}) then has the asymptotic scaling
\begin{equation}
    \mathcal{P}_h^\delta(k \ll k_p) \propto k^2 \,,
\end{equation}
which can also be seen in the plot of Fig.~\ref{fig:tensorPk}. 
Thus we verify our earlier claim that the IR tail of the tensor power spectrum scales like $k^2$ in the delta function case. This is of course unphysical; causality\footnote{See Ref.~\cite{Cai:2019} for a discussion of the origin of this $\sim k^3$ super-horizon scaling, as well as physical conditions under which it can be violated.} dictates that the super-horizon tail should fall off no slower than $k^3$ for a bounded source. For this reason when we actually go about computing the B-mode signal from induced tensor perturbations, we will use a different ansatz for the scalar source.

\subsubsection{Boxcar Function}

Consider now the case of a ``boxcar function'' scalar power spectrum, 
\begin{equation}\label{eq:PRbox}
    \mathcal{P}_\mathcal{R}^B(k) = \mathcal{A} \, B(k,k_p,\Delta) \,,
\end{equation}
where $\mathcal{A}$ is the amplitude of the peak and $B(k,k_p,\Delta)$ is the normalized boxcar function, which is zero everywhere except for the interval $k_p e^{-\Delta/2} \leq k \leq k_p e^{\Delta/2}$ where it is equal to $1/\Delta$, i.e.
\begin{equation}
    B(k,k_p,\Delta) = \begin{cases} \frac{1}{\Delta} & k_p e^{- \Delta/2} \leq k \leq k_p e^{\Delta/2} \,, \\ 0 & \text{otherwise} \,. \end{cases}
\end{equation}
This function approximates a delta function in the limit of small width $\Delta \rightarrow 0$, and one can check that integrating $B(k,k_p,\Delta)$ over all $k$ yields $k_p$ in this limit, just as integrating $\delta(\ln(k/k_p))$ over all $k$ yields $k_p$. Taking this ansatz for the scalar power spectrum is particularly convenient since the calculation of $\mathcal{P}_h$ remains semi-analytically tractable. In contrast to the delta function, though, the spectrum we find will demonstrate the proper $\sim k^3$ causal scaling in the IR. 

Converting to the variables $u$ and $v$ and inserting the ansatz of Eq.~(\ref{eq:PRbox}) into the tensor power spectrum Eq.~(\ref{eq:Phuv}) yields
\begin{equation}
    \mathcal{P}_h^B(x, k) = \frac{\mathcal{A}^2}{2} \! \int_0^\infty \!\! dv \, \int_{|1-v|}^{1+v} \!\! du \left( \! \frac{4v^2 - (1 + v^2 - u^2)^2}{u v} \! \right)^2 \, \tilde{\mathcal{I}}_{\rm RD}(x,u,v)^2 \, B\! \left( \! u, \frac{k}{k_p}, \Delta \! \right) B \!\left(\! v, \frac{k}{k_p}, \Delta \! \right) \,,
\end{equation}
where the boxcar function in these new variables is
\begin{equation}
    B(v, k/k_p, \Delta) = \begin{cases} \frac{1}{\Delta} & \frac{k_p}{k} e^{-\Delta/2} \leq v \leq \frac{k_p}{k} e^{\Delta/2} \,, \\ 0 & \text{otherwise} \,. \end{cases}
\end{equation}
For the variable $v$, it is trivial to use the boxcar function to change the integration limits, replacing $[0,\infty) \rightarrow [\frac{k_p}{k} e^{-\Delta/2}, \frac{k_p}{k} e^{\Delta/2}]$ and $B(v, k/k_p, \Delta) \rightarrow 1/\Delta$. For the variable $u$, the new integration limits are a little more subtle. Defining
\begin{equation}
    u_{\rm min} \equiv \text{max} \left( |1-v|, \frac{k_p}{k} e^{-\Delta/2} \right) \,, \,\,\,\,\,\, u_{\rm max} \equiv \text{min}\left(1+v, \frac{k_p}{k} e^{\Delta/2} \right) \,,
\end{equation}
we can now use the boxcar function to replace $[|1-v|, 1+v] \rightarrow [u_{\rm min}, u_{\rm max}]$ and $B(u, k/k_p, \Delta) \rightarrow 1/\Delta$. The tensor power spectrum becomes
\begin{equation}\label{eq:Phbox}
    \mathcal{P}_h^B(x, k) = \frac{\mathcal{A}^2}{2 \Delta^2} \! \int_{\frac{k_p}{k} e^{-\Delta/2}}^{\frac{k_p}{k} e^{\Delta/2}} \!\! dv \, \int_{u_{\rm min}}^{u_{\rm max}} \!\! du \left( \! \frac{4v^2 - (1 + v^2 - u^2)^2}{u v} \! \right)^2 \, \tilde{\mathcal{I}}_{\rm RD}(x,u,v)^2 \,.
\end{equation}
At this point we need to evaluate the expression numerically. Thankfully, the limited integration range makes this process quite fast and accurate. The resulting tensor power spectrum is shown in Fig.~\ref{fig:tensorPk} for two sample peak locations. We see that at large $k$, the spectrum in the boxcar case very closely resembles the delta function spectrum. In particular, momentum conservation still imposes a cutoff at $k = 2 k_p$. The peak of the tensor spectrum is at slightly smaller $k$ compared with the location of the scalar peak $k_p$. As one goes to smaller $k$, the spectrum initially falls off as $k^2$ before transitioning into the characteristic $k^3$ scaling expected of a causal source. It is very important that in calculating the B-mode spectrum we use a physical ansatz for $\mathcal{P}_\mathcal{R}(k)$ like the boxcar function or a lognormal function, as opposed with the unphysical delta function. CMB observables are sensitive to long-wavelength modes, and in the IR the delta function spectrum features an artificial enhancement. Finally, we note that damping effects~\cite{Domenech:2025bvr} can lead to slight modifications in the scaling behavior of the low frequency tail. While incorporating these effects lies beyond the scope of this work, it may become important in the context of future high-precision B-mode polarization measurements. 

\section{B-Mode Polarization}\label{sec:Bmode}

The induced background of GWs sourced by enhanced small scale curvature perturbations has been exhaustively studied in the literature (see e.g.~\cite{Ananda2007,Baumann2007,Kohri:2018,Domenech:2021}). What has not been appreciated before, though, is that these same tensor perturbations can also generate a B-mode polarization signal in the CMB. In this section, we compute this signal. We begin in \ref{sec:Bmodereview} by reviewing the formalism for calculating the angular spectrum of B-mode polarization $C_\ell^{BB}$ given a form for the tensor power spectrum $\mathcal{P}_h(k)$. Sec.~\ref{sec:whentoevaluate} provides details of our numerical implementation for this particular source. Finally, Sec.~\ref{sec:results} presents resulting sample spectra.

\subsection{CMB Polarization: Theory}\label{sec:Bmodereview}

CMB polarization arises indirectly from temperature anisotropies sourced by primordial metric perturbations, which are converted into polarization via Thomson scattering around the era of recombination. Perturbations in the early universe --- including both scalar and tensor --- induce anisotropies in the temperature distribution of photons. Around the onset of recombination, when photons begin to free stream, quadrupole anisotropies can develop. When incoming radiation with a quadrupole temperature anisotropy Thomson scatters off free electrons, it generates net linear polarization in the outgoing radiation. As the number density of free electrons drops sharply afterwards, this linear polarization is imprinted on the CMB. 

The resulting polarization can be decomposed into two types: the E-mode pattern, which is gradient-like, curl-free, and parity even; and the B-mode pattern, which is curl-like, divergence-free, and parity odd. Only tensor-induced quadrupoles result in the latter pattern, since they break axial symmetry. Hence the origin of the standard lore that an observation of B-mode polarization (above astrophysical foregrounds) would constitute smoking gun evidence for primordial tensor perturbations.\footnote{Gravitational lensing of E-mode polarization can also produce the B-mode pattern, so ascertaining whether or not a measured B-mode signal is truly primordial requires appropriate delensing. One must also contend with foreground contamination in the form of polarized synchrotron radiation and thermal dust emitted from the Milky Way.} 

We now turn to make this quantitative. Consider photons traveling along the direction $\mathbf{\hat{n}'}$ to position $\mathbf{x}$, where they scatter for the last time at time $\eta$. We denote by $\delta T/T(\eta, \mathbf{x}; \mathbf{n}')$ their temperature anisotropy, which takes the form~\cite{Rubakov:2011}
\begin{equation}\label{eq:deltaToverTpos}
    \frac{\delta T}{T}(\eta,\mathbf{x};\mathbf{n'}) = \frac{1}{2} \int_0^\eta d\eta_1 V(\eta,\eta_1) \int d^3 k \int_{\eta_1}^\eta d\eta_2 \, n_i' \frac{\partial h_{ij}}{\partial \eta_2} n_j' e^{i \mathbf{k} \cdot (\mathbf{x}-\mathbf{n'}(\eta - \eta_2))} \,,
\end{equation}
where $V(\eta,\eta_1)$ is the visibility function
\begin{equation}
    V(\eta, \eta_1) = e^{-\kappa(\eta,\eta_1)} \frac{d\kappa}{d\eta_1} \,,
\end{equation}
defined in terms of the optical depth $\kappa(\eta,\eta_1)$,
\begin{equation}
    \kappa(\eta,\eta_1) = \int_{\eta_1}^{\eta} d\tilde{\eta} \, a(\tilde{\eta}) \sigma_T n_e(\tilde{\eta}) \,,
\end{equation}
with $n_e$ the electron number density and $\sigma_T$ the Thomson cross section. The innermost integral in this formula --- which represents the time variation of the tensor perturbation along a photon geodesic, integrated over the line of sight --- captures the contribution from the integrated Sachs-Wolfe (ISW) effect, which is essentially the only source of temperature anisotropy from tensor perturbations. The result is then integrated over all times $\eta_1$ that the second-to-last scattering could have occurred, weighted by the visibility function. 

Letting $\mathbf{n}$ be the direction of observation and $\eta_0$ the conformal time today, we can write the point of last scattering as $\mathbf{x} = (\eta_0 - \eta) \mathbf{n}$. Substituting this into Eq.~(\ref{eq:deltaToverTpos}) and focusing on a single Fourier mode gives
\begin{equation}\label{eq:deltaToverTk}
    \frac{\delta T}{T}(\eta,\mathbf{k};\mathbf{n},\mathbf{n'}) = \frac{1}{2} e^{i \mathbf{k} \cdot \mathbf{n} (\eta_0 - \eta)} \int_0^\eta d\eta_1 V(\eta,\eta_1) \int_{\eta_1}^\eta d\eta_2 \, n_i' \frac{\partial h_{ij}}{\partial \eta_2} n_j' e^{-i\mathbf{k} \cdot \mathbf{n'}(\eta - \eta_2)} \,.
\end{equation}
By integrating this expression over arrival directions $\mathbf{n'}$ as well as times that last scattering could occur $\eta$, weighted by the visibility function, one obtains for the CMB polarization tensor~\cite{Rubakov:2011}
\begin{equation}\label{eq:Pab}
    P_{ab}(\mathbf{k}; \mathbf{n}) = \frac{3}{4\pi} \int d^2 n' \, \left( \frac{1}{2} \left( 1 - (\mathbf{n} \cdot \mathbf{n'})^2 \right) g_{ab} - (\mathbf{n'} \cdot \mathbf{\hat{e}}_a) (\mathbf{n'} \cdot \mathbf{\hat{e}}_b) \right) \int_0^{\eta_0} d\eta \, V(\eta_0,\eta) \frac{\delta T}{T}(\eta,\mathbf{k};\mathbf{n},\mathbf{n'}) \,,
\end{equation}
where $g_{ab}$ is the 2-dimensional metric on the celestial sphere and $\mathbf{\hat{e}}_{a}$ are a set of orthonormal basis vectors. The tensor structure in parentheses projects out the transverse, traceless part of the quadrupole moment\footnote{The requirement of a quadrupole anisotropy is reflected in the fact that $\mathbf{n'}$ enters in a bilinear combination.} of the temperature anisotropy onto the plane perpendicular to $\mathbf{n}$. 

In general, one can decompose the CMB polarization tensor into E- and B-mode parts,
\begin{equation}
    P_{ab}(\mathbf{k}; \mathbf{n}) = - \sum_{\ell=2}^\infty \sum_{m = - \ell}^\ell \bigg( a_{\ell m}^{E}(\mathbf{k}) [Y_{\ell m}^{(E)}]_{ab}(\mathbf{n}) + a_{\ell m}^{B}(\mathbf{k}) [Y_{\ell m}^{(B)}]_{ab}(\mathbf{n}) \bigg) \,,
\end{equation}
where $[Y_{\ell m}^{(E,B)}]_{ab}(\mathbf{n})$ are the E- and B-mode tensor spherical harmonics, which constitute a complete orthonormal system for second-rank symmetric traceless tensors on the sphere. The latter is related to the usual spherical harmonics via
\begin{equation}
    [Y_{\ell m}^{(B)}]_{ab} = \sqrt{\frac{(\ell-2)!}{2(\ell+2)!}} \bigg( \epsilon_b^c \, \nabla_a \nabla_c Y_{\ell m} + \epsilon_a^c \, \nabla_c \nabla_b Y_{\ell m} \bigg) \,.
\end{equation}
To isolate the coefficients $a_{\ell m}^B(\mathbf{k})$, one can invert
\begin{equation}\label{eq:almdef}
    a_{\ell m}^B(\mathbf{k}) = - \int d^2 n \, [Y_{\ell m}^{(B)}]_{ab}^*(\mathbf{n}) P^{ab}(\mathbf{k}; \mathbf{n}) \,.
\end{equation}
The angular spectrum of B-mode polarization is defined in terms of these coefficients as
\begin{equation}\label{eq:ClBBdef}
    C_\ell^{BB} = \frac{1}{2\ell+1} \sum_m \int \frac{d^3k}{(2\pi)^3} \langle a_{\ell m}^B(\mathbf{k}) a_{\ell m}^B(\mathbf{k})^* \rangle \,.
\end{equation}

To actually evaluate these expressions, we work with spherical coordinates in a frame with the azimuthal axis directed along $\mathbf{k}$. The metric is $ds^2 = g_{ab} dx^a dx^b = d\theta^2 + \sin^2 \theta d\phi^2$, the Cartesian components of the basis vectors $\mathbf{\hat{e}}_{a = \theta, \phi}$ are
\begin{equation}
    \mathbf{\hat{e}}_\theta = ( - \cos \theta \cos \phi, \, - \cos \theta \sin \phi, \, \sin \theta) \,, \,\,\,\,\,\, \mathbf{\hat{e}}_\phi = (\sin \theta \sin \phi , \, - \sin \theta \cos \phi, \, 0) \,,
\end{equation}
and the components of $\mathbf{n}$ and $\mathbf{n'}$ are
\begin{equation}
    \mathbf{n} = (\sin \theta \cos \phi, \, \sin \theta \sin \phi, \, \cos \theta) \,, \,\,\,\,\,\, \mathbf{n'} = (\sin \theta' \cos \phi', \, \sin \theta' \sin \phi', \, \cos \theta') \,.
\end{equation}
Consider first the polarization tensor of Eq.~(\ref{eq:Pab}), with $\delta T/T$ in Eq.~(\ref{eq:deltaToverTk}). These equations feature the combination
\begin{equation}
    n_i' h_{ij} n_j' = \sum_{\lambda = +,\times} n_i' \varepsilon_{ij}^\lambda n_j' h_\lambda \,,
\end{equation}
where we have decomposed $h_{ij}$ into the two polarizations $\lambda = +,\times$ and sum over the contribution from both. The convolutions of $\varepsilon_{ij}^\lambda$ with the components $n_i'$ in this frame are
\begin{equation}
    n_i' \varepsilon_{ij}^+ n_j' = \sin^2 \theta' \cos 2\phi' \,, \,\,\,\,\,\, n_i' \varepsilon_{ij}^\times n_j' = \sin^2 \theta' \sin 2\phi' \,.
\end{equation}
Upon integrating over $\phi'$, the only surviving projections will then be of the form
\begin{subequations}
\begin{equation}
    \left[ \frac{1}{2} \left( 1 - (\mathbf{n} \cdot \mathbf{n'})^2 \right) g_{\theta\theta} - (\mathbf{n'} \cdot \mathbf{\hat{e}}_\theta) (\mathbf{n'} \cdot \mathbf{\hat{e}}_\theta) \right] \rightarrow - \frac{1}{4} (1+\cos^2 \theta) \sin^2 \theta' \cos 2(\phi - \phi') \,,
\end{equation}
\begin{equation}
    \left[ \frac{1}{2} \left( 1 - (\mathbf{n} \cdot \mathbf{n'})^2 \right) g_{\phi \phi} - (\mathbf{n'} \cdot \mathbf{\hat{e}}_\phi) (\mathbf{n'} \cdot \mathbf{\hat{e}}_\phi) \right] \rightarrow \frac{1}{4} \sin^2 \theta (1+\cos^2 \theta) \sin^2 \theta' \cos 2(\phi - \phi') \,,
\end{equation}
\begin{equation}
    \left[ \frac{1}{2} \left( 1 - (\mathbf{n} \cdot \mathbf{n'})^2 \right) g_{\theta \phi} - (\mathbf{n'} \cdot \mathbf{\hat{e}}_\theta) (\mathbf{n'} \cdot \mathbf{\hat{e}}_\phi) \right] \rightarrow \frac{1}{2} \sin \theta \cos \theta \sin^2 \theta' \sin 2(\phi - \phi') \,.
\end{equation}
\end{subequations}
Using these expressions, one can perform the angular integrals over $\mathbf{n'}$ to obtain $P_{ab}(\mathbf{k}; \mathbf{n})$. The result enters into the coefficients $a_{\ell m}^B(\mathbf{k})$ of Eq.~(\ref{eq:almdef}). To perform the angular integrals over $\mathbf{n}$, we note that the B-mode tensor spherical harmonics $Y_{\ell m}^{(B)}(\mathbf{n})$ have the following components in this coordinate system
\begin{subequations}
\begin{equation}
    [Y_{\ell m}^{(B)}]_{\theta \theta}(\mathbf{n}) = - \sqrt{\frac{(\ell-2)!}{2(\ell+2)!}} \frac{2im}{\sin \theta} \left( \frac{\partial}{\partial \theta} - \cot \theta \right) Y_{\ell m}(\mathbf{n}) \,,
\end{equation}
\begin{equation}
    [Y_{\ell m}^{(B)}]_{\phi \phi}(\mathbf{n}) = \sqrt{\frac{(\ell-2)!}{2(\ell+2)!}}\, 2im \, \sin \theta \left( \frac{\partial}{\partial \theta} - \cot \theta \right) Y_{\ell m}(\mathbf{n}) \,,
\end{equation}
\begin{equation}
    [Y_{\ell m}^{(B)}]_{\theta \phi}(\mathbf{n}) = \sqrt{\frac{(\ell-2)!}{2(\ell+2)!}} \sin \theta \left( 2 \frac{\partial^2}{\partial \theta^2} + \ell(\ell+1) \right) Y_{\ell m}(\mathbf{n}) \,.
\end{equation}
\end{subequations}
Since the azimuthal angle appears in the integral of Eq.~(\ref{eq:almdef}) only through the factors $\cos 2 \phi$ and $\sin 2 \phi$, the integral is non-vanishing only for the spherical harmonics with $m = \pm 2$. Thus the only non-vanishing coefficients will be $a_{\ell, \pm 2}^B$. 

As a final step before using these coefficients to evaluate $C_\ell^{BB}$ in Eq.~(\ref{eq:ClBBdef}), we decompose the tensor perturbation as
\begin{equation}
    h_\lambda(\eta, \mathbf{k}) = h_{\mathbf{k}}^\lambda \, \mathcal{T}(\eta,k) \,,
\end{equation}
where $h_{\mathbf{k}}^\lambda \equiv h_\lambda(\eta_i, \mathbf{k})$ is the ``initial'' amplitude and $\mathcal{T}(\eta,k)$ is the transfer function. The statistical properties of the initial amplitudes are encoded in the dimensionless tensor power spectrum $\mathcal{P}_h(k)$ defined in Eq.~(\ref{eq:Phdef}). In practice, we will choose horizon crossing for each mode as the time to set these initial conditions since, as we will demonstrate shortly, it is here that the tensor perturbations peak. Meanwhile, a mode function's subsequent sub-horizon evolution is captured by the transfer function $\mathcal{T}(\eta,k)$, which in the absence of anisotropic stress obeys the sourceless wave equation\footnote{In reality, there is some small amount of anisotropic stress from free streaming neutrinos. We omit the term here, but account for this effect in our numerical implementation by using the transfer functions computed by the Boltzmann solver \texttt{CLASS}~\cite{CLASS:I,CLASS:II}.}
\begin{equation}
    \mathcal{T}'' + 2 \mathcal{H}\mathcal{T}' + k^2 \mathcal{T} \simeq 0 \,.
\end{equation}
We comment that in performing this decomposition, we are making an approximation, since the actual evolution of the mode function of Eq.~(\ref{eq:hsol}) post-horizon re-entry does not exactly coincide with the evolution of the transfer function. However, we have verified numerically that the behavior fairly faithfully approximates the true sub-horizon evolution, in particular for those wavenumbers $k < 1 \, \text{Mpc}^{-1}$ which give the dominant contribution to the B-mode signal. If anything, the B-mode estimates we present are conservative, since the transfer function decays slightly more abruptly as compared with the actual mode function evolution.

\begin{figure}[h!]
\centering
\includegraphics[width=0.65\textwidth]{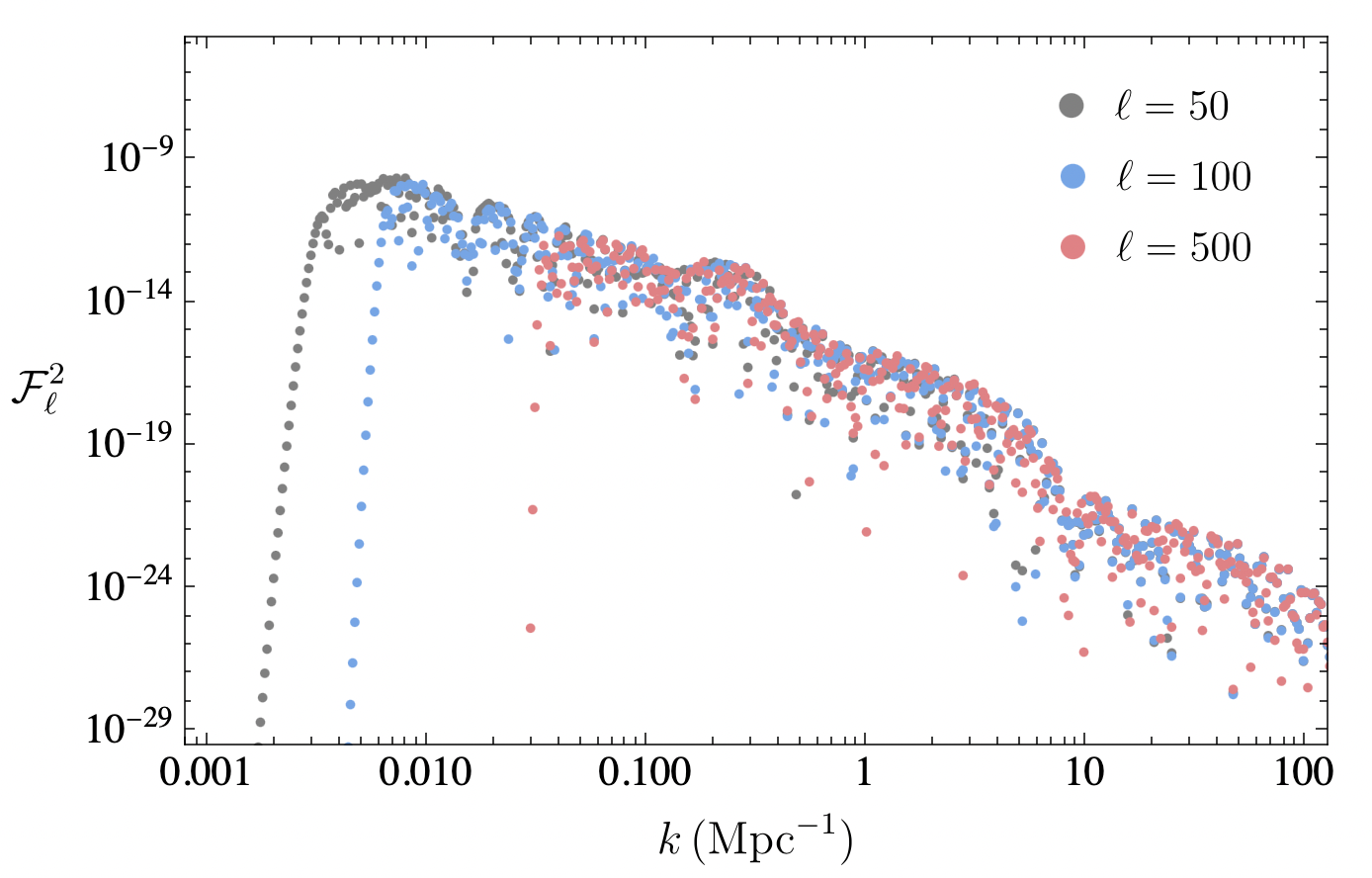}
\caption{Window function of Eq.~(\ref{eq:Fldef}) squared, $\mathcal{F}_\ell(k)^2$, for three sample multipoles: $\ell = 50$ (gray), $\ell = 100$ (light blue), and $\ell = 500$ (light red). The shape of this function, when convolved with the tensor power spectrum $\mathcal{P}_h$, determines the shape of the angular spectrum of B-mode polarization according to Eq.~(\ref{eq:ClBB}).}
\label{fig:Flsq}
\end{figure}

Combining these results, carrying out the integrals explicitly, and summing over polarizations, we find the angular spectrum of B-mode polarization Eq.~(\ref{eq:ClBBdef}) can be written
\begin{equation}\label{eq:ClBB}
    C_\ell^{BB} = 36 \pi \int_0^\infty \frac{dk}{k} \mathcal{P}_h(k) \mathcal{F}_\ell(k)^2 \,,
\end{equation}
where we have defined the function
\begin{equation}\label{eq:Fldef}
\begin{split}
    \mathcal{F}_\ell(k) = \int_0^{\eta_0} d\eta \, V(\eta_0,&\eta) \, \bigg[ \frac{\ell+ 2}{2 \ell + 1} j_{\ell - 1} \left[ k(\eta_0 - \eta) \right] - \frac{\ell - 1}{2\ell + 1} j_{\ell + 1} \left[ k (\eta_0 - \eta) \right] \bigg] \\
    & \times \int_0^\eta d\eta_1 \, V(\eta,\eta_1) \int_{\eta_1}^{\eta} d\eta_2  \, \frac{j_2[k(\eta - \eta_2)]}{k^2 (\eta - \eta_2)^2} 
  \left(\! \frac{\partial \mathcal{T}(\eta_2,k)}{\partial \eta_2} \!\right) \,,
\end{split}
\end{equation}
with $j_n(x)$ a spherical Bessel function of the first kind. The function $\mathcal{F}_\ell(k)$ is entirely determined by the background evolution and should be thought of as a kind of window function which, when integrated against the power spectrum, picks out those modes to which the CMB is most sensitive. We show $\mathcal{F}_\ell(k)^2$ for three sample multipoles in Fig.~\ref{fig:Flsq}.
%

\subsection{Numerical Implementation}\label{sec:whentoevaluate}

The analytical formulas derived in the previous section are useful for developing intuition, but difficult to evaluate numerically due to the nested structure and highly oscillatory nature of the integrands. To minimize numerical error, we use the Boltzmann solver \texttt{CLASS} for our implementation~\cite{CLASS:I,CLASS:II}. This also allows us to use the exact visibility function and the exact tensor mode transfer function. We utilize the \texttt{external-pk} module with the induced tensor power spectra of Sec.~\ref{sec:theory} as input. Given these initial conditions, \texttt{CLASS} uses the Einstein-Boltzmann equations and polarization source functions to propagate the signal forward and evaluate the line-of-sight and Fourier integrals to obtain the angular spectrum of B-mode polarization $C_\ell^{BB}$, as described in Sec.~\ref{sec:Bmodereview}. 

\begin{figure}[h!]
\centering
\includegraphics[width=0.65\textwidth]{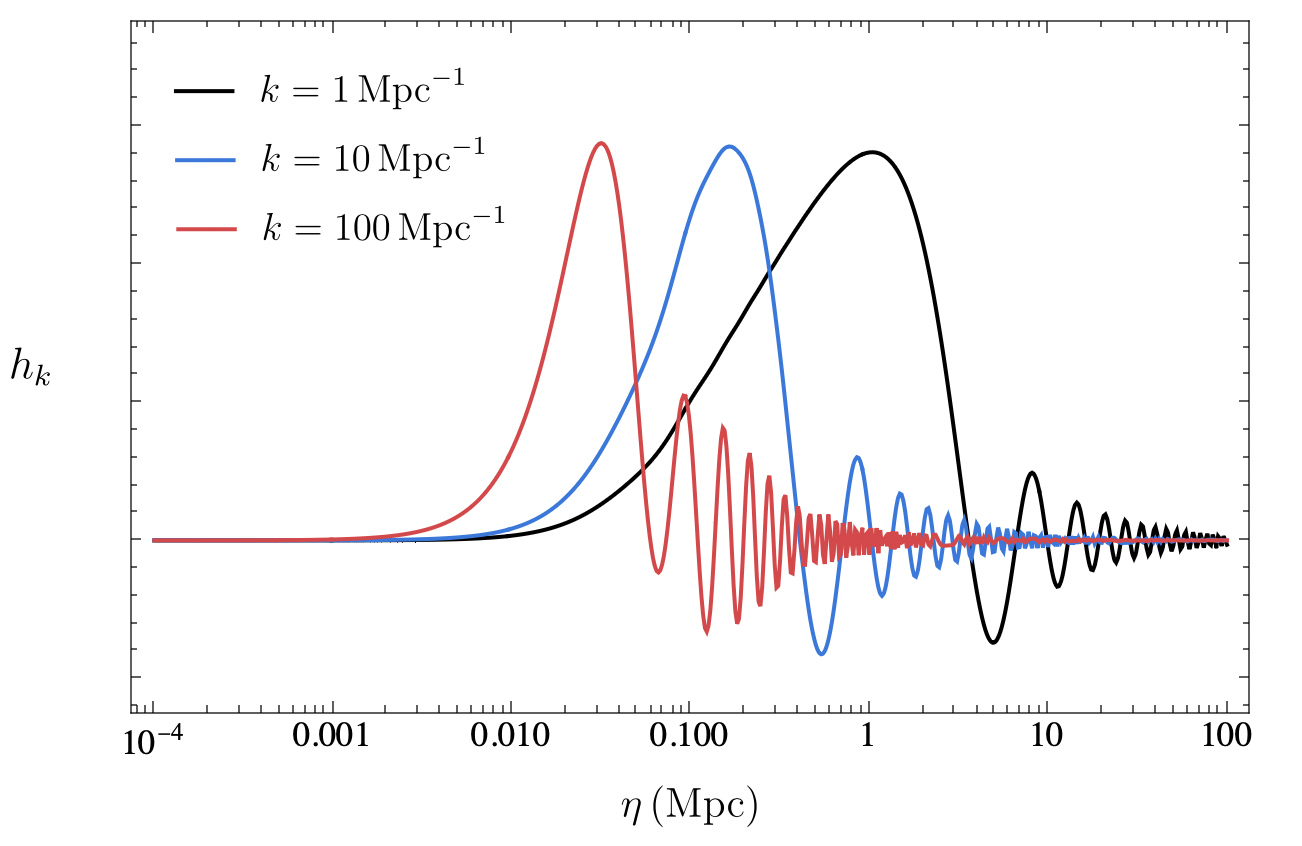}
\caption{Rescaled tensor mode $h_{\mathbf{k}}(\eta)$ of Eq.~(\ref{eq:hk}) as a function of time for three sample wavenumbers. The peak location is fixed as $k_p = 100 \, \text{Mpc}^{-1}$. We observe that the maximum amplitude is achieved within an order 1 factor of horizon crossing $\eta = 1/k$ for each mode.}
\label{fig:hevo}
\end{figure}
It is essential that we input the initial tensor power spectrum evaluated at a time once the modes have finished growing, but before horizon entry and subsequent decay. A natural choice is to define the initial power spectrum at horizon crossing for each mode $\eta = 1/k$, or equivalently $x=1$, 
\begin{equation}
    \mathcal{P}_h^{\rm ini}(k) \equiv \mathcal{P}_h(x=1, k) \,.
\end{equation}
To verify that this is a reasonable choice, we can solve the wave equation of Eq.~(\ref{eq:tensor_eom}) for $h_{\mathbf{k}}(\eta)$ and study the behavior as a function of time. For simplicity, suppose that all power is localized in one mode $k_p$, corresponding to the curvature perturbation\footnote{Note that this expression must have some azimuthal dependence in order to break spherical symmetry, such that we can source tensor modes. This can be seen from the fact that the integral over $\phi$ in the definition of $S_{\mathbf{k}}$ vanishes unless $\mathcal{R}(\mathbf{q})$ has some angular dependence.}
\begin{equation}
    \mathcal{R}_{\mathbf{k}} \sim k_p\, \delta(k - k_p) \,.
\end{equation}
For this form of $\mathcal{R}$, one can use the delta functions to perform the integrals over $q$ and $\cos \theta$ in the source term of Eq.~(\ref{eq:source_term}), yielding 
\begin{equation}
    S_{\mathbf{k}}(\eta) \sim (4 k_p^2 - k^2) f(\eta, k_p, k_p) \Theta(2k_p - k) \,.
\end{equation}
Substituting this source into Eq.~(\ref{eq:hsol}) we find up to constants that
\begin{equation}\label{eq:hk}
    h_{\mathbf{k}}(\eta) \sim \left( \frac{4 k_p^2 - k^2}{k \eta} \right) \Theta(2 k_p - k) \int_{0}^\eta d \bar{\eta} \, \sin[k(\eta - \bar{\eta})] \bar{\eta} f(\bar{\eta}, k_p, k_p) \,.
\end{equation}
We evaluate the integral over $\bar{\eta}$ numerically and plot the result in Fig.~\ref{fig:hevo} for three sample values of $k < k_p$. We observe that the modes achieve their maximum values shortly (i.e. within an $\mathcal{O}(1)$ factor) after horizon crossing. This motivates us to define the initial spectrum at $\eta = 1/k$ for each mode. We comment that this is a conservative choice; more generally, one should solve for the time at which $h_k$ is maximal for each wavenumber $k$. In practice since a majority of the signal will come from small $k \lesssim 10\, \text{Mpc}^{-1}$ (see e.g. Fig.~\ref{fig:spectralshape}), for which $\eta = 1/k$ is a very good approximation, this procedure incurs minimal error.

\subsection{Numerical Results}\label{sec:results}

\begin{figure}[b!]
\centering
\includegraphics[width=0.7\textwidth]{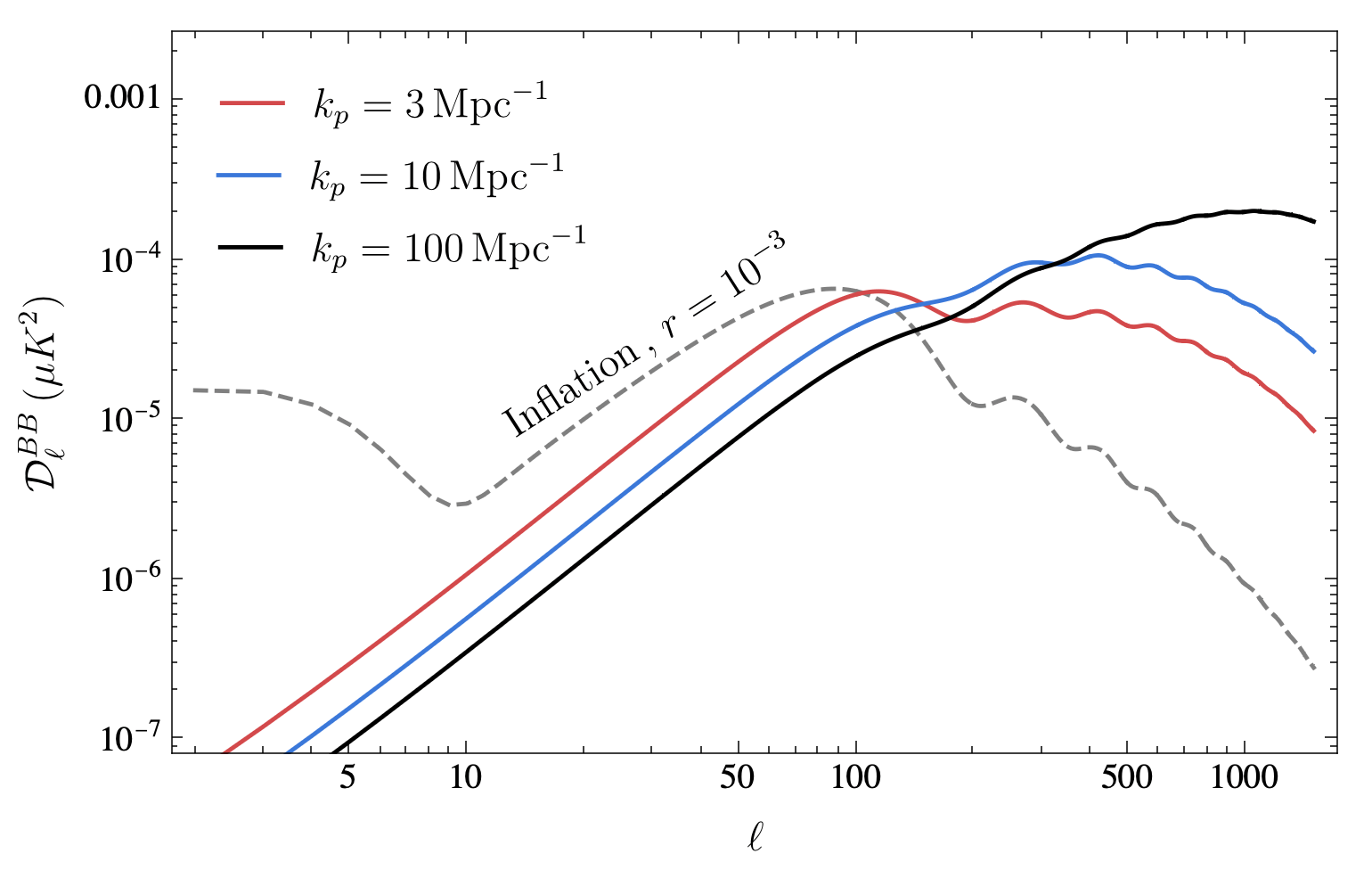}
\hspace{3mm}
\includegraphics[width=0.7\textwidth]{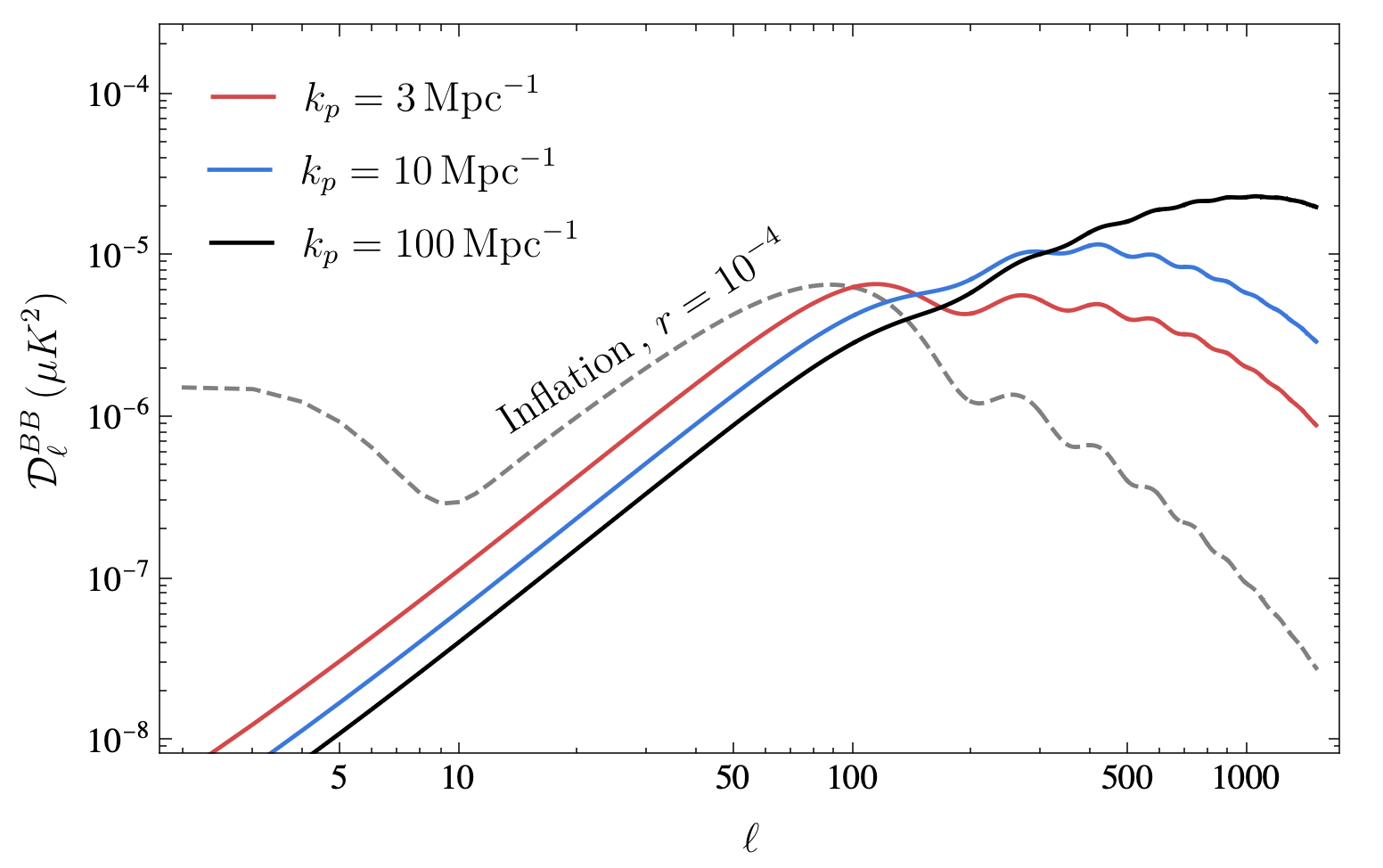}
\caption{Angular spectrum of B-mode polarization $\mathcal{D}_\ell^{BB}$ for induced tensor modes as compared with the inflationary signal (gray, dashed) for tensor-to-scalar ratios of $r = 10^{-3}$ (top) and $r = 10^{-4}$ (bottom). We model the scalar power spectra with the boxcar function of Eq.~(\ref{eq:PRbox}), leading to the tensor power spectrum in Eq.~(\ref{eq:Phbox}). We presume a narrow peak $\Delta = 10^{-2}$ centered on wavenumbers $k_p = 3\, \text{Mpc}^{-1}$ (red), $10\, \text{Mpc}^{-1}$ (blue), and $100\, \text{Mpc}^{-1}$ (black). In all cases, the amplitude is chosen such that the signal-to-noise ratio matches the respective value from the inflationary signal. For computing S/N, we suppose noise spectra for a future CMB-S4 like experiment, taking $\Delta_p = 1.5 \, \mu\text{K} \cdot \text{arcmin}$ as the polarization noise level in the maps, $\Theta_{\rm FWHM} = 1 \, \text{arcmin}$ as the FWHM of the Gaussian beam, $\ell_{\rm min} = 30$ as the minimum multipole in the sum of Eq.~(\ref{eq:SNdef}), $\ell_{\rm max} = 3000$ as the maximum multipole, and $\ell_{\rm knee} = 60$ and $\gamma = -3$ as the fitting parameters in Eq.~(\ref{eq:NlBB}), in accordance with CMB-S4 forecasts~\cite{CMBS4:2020}.}
\label{fig:DlBB}
\end{figure}
In Fig.~\ref{fig:DlBB} we show the angular spectrum of B-mode polarization plotted in terms of $\mathcal{D}_\ell^{BB}$, related to the $C_\ell^{BB}$ of Eq.~(\ref{eq:ClBB}) as
\begin{equation}\label{eq:DlBB}
    \mathcal{D}_\ell^{BB} \equiv \frac{\ell(\ell+1)}{2\pi} T_0^2 C_\ell^{BB} \,.
\end{equation}
The top panel shows spectra for a narrow peak located at $k_p = 3 \, \text{Mpc}^{-1}$ (red), $10 \, \text{Mpc}^{-1}$ (blue), and $100 \, \text{Mpc}^{-1}$ (black) as compared with the an inflationary spectrum with tensor-to-scalar ratio $r = 10^{-3}$. Concretely, we model the scalar power spectra $\mathcal{P}_\mathcal{R}$ by the boxcar function of Eq.~(\ref{eq:PRbox}) with width $\Delta = 10^{-2}$, leading to the tensor power spectrum of Eq.~(\ref{eq:Phbox}). The amplitudes are chosen such that the signal-to-noise (S/N) ratio for each benchmark matches the signal-to-noise ratio of the inflationary signal with $r = 10^{-3}$, whose square is computed as
\begin{equation}\label{eq:SNdef}
    (\text{S}/\text{N})^2 = \sum_{\ell = \ell_{\rm min}}^{\ell_{\rm max}} \frac{(2\ell+1) f_{\rm sky}}{2} \left( \frac{\mathcal{D}_\ell^{BB}}{\mathcal{D}_\ell^{BB} + \mathcal{N}_\ell^{BB}} \right)^2 \,,
\end{equation}
where $\mathcal{N}_\ell^{BB}$ is the noise power spectrum and $f_{\rm sky}$ is the observed sky fraction. We parameterize the projected noise spectra as
\begin{equation}\label{eq:NlBB}
    \mathcal{N}_\ell^{BB} = \frac{\ell(\ell+1)}{2\pi} N_{\rm white} \bigg( 1 + \left( \frac{\ell}{\ell_{\rm knee}} \right)^{\gamma} \bigg) \,,
\end{equation}
where the white noise term is
\begin{equation}
    N_{\rm white} = \Delta_p^2 \exp \left[ \frac{\ell(\ell+1)}{8 \ln 2} \Theta_{\rm FWHM}^2 \right] \,,
\end{equation}
and adopt values from Ref.~\cite{CMBS4:2020} corresponding to forecasts for CMB-S4 (see also Ref.~\cite{CMBS4}). In particular, we take $\Delta_p = 1.5 \, \mu\text{K} \cdot \text{arcmin}$ for the polarization noise level and $\Theta_{\rm FWHM} = 1-30 \, \text{arcmin}$ for the full-width half-maximum (FWHM) of the Gaussian beam, corresponding roughly to the planned small aperature telescope (SAT) survey from CMB-S4. The projections of Ref.~\cite{CMBS4:2020} find $\ell_{\rm knee} = 50-60$ and $\gamma = -2$ to $-3$ for small aperture data, and so we fix $\ell_{\rm knee} = 60$ and $\gamma = -3$ for $\Theta_{\rm FHWM} = 1 \,\text{arcmin}$ when computing S/N in the curves of Fig.~\ref{fig:DlBB}. Finally, we take $\ell_{\rm min} = 30$, corresponding to the largest scales accessible to ground-based surveys, and $\ell_{\rm max} = 3000$, which is consistent with the instrument's angular resolution for small $\Theta_{\rm FWHM}$.

We observe that the induced tensor mode spectra generically peak on smaller scales (larger $\ell$) as compared with the inflationary case. Correspondingly, larger amplitudes are required to achieve the same S/N as those spectra peaking at smaller $\ell$.
\begin{figure}[h!]
\centering
\includegraphics[width=0.6\textwidth]{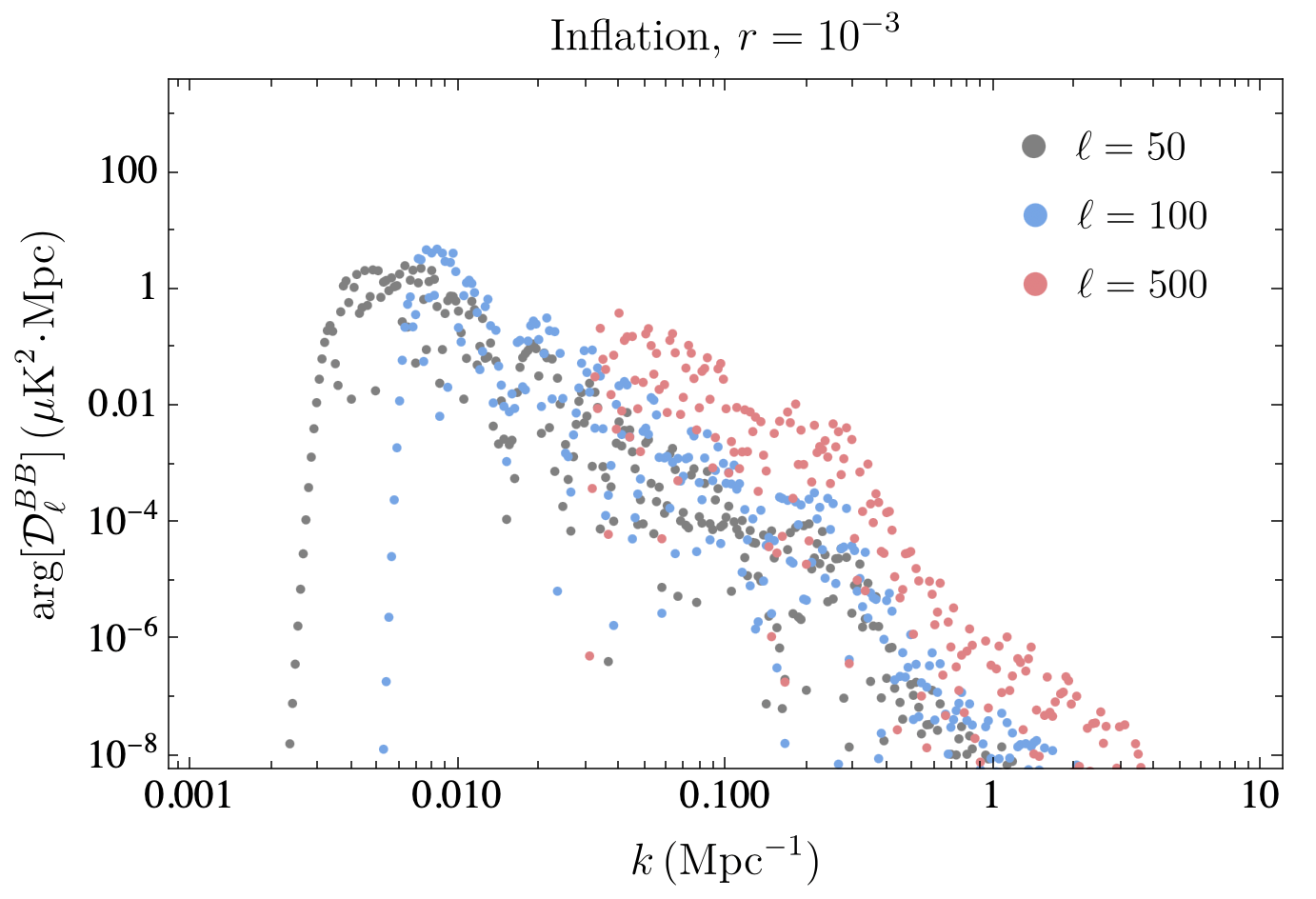}\\
\hspace{3mm}
\includegraphics[width=0.6\textwidth]{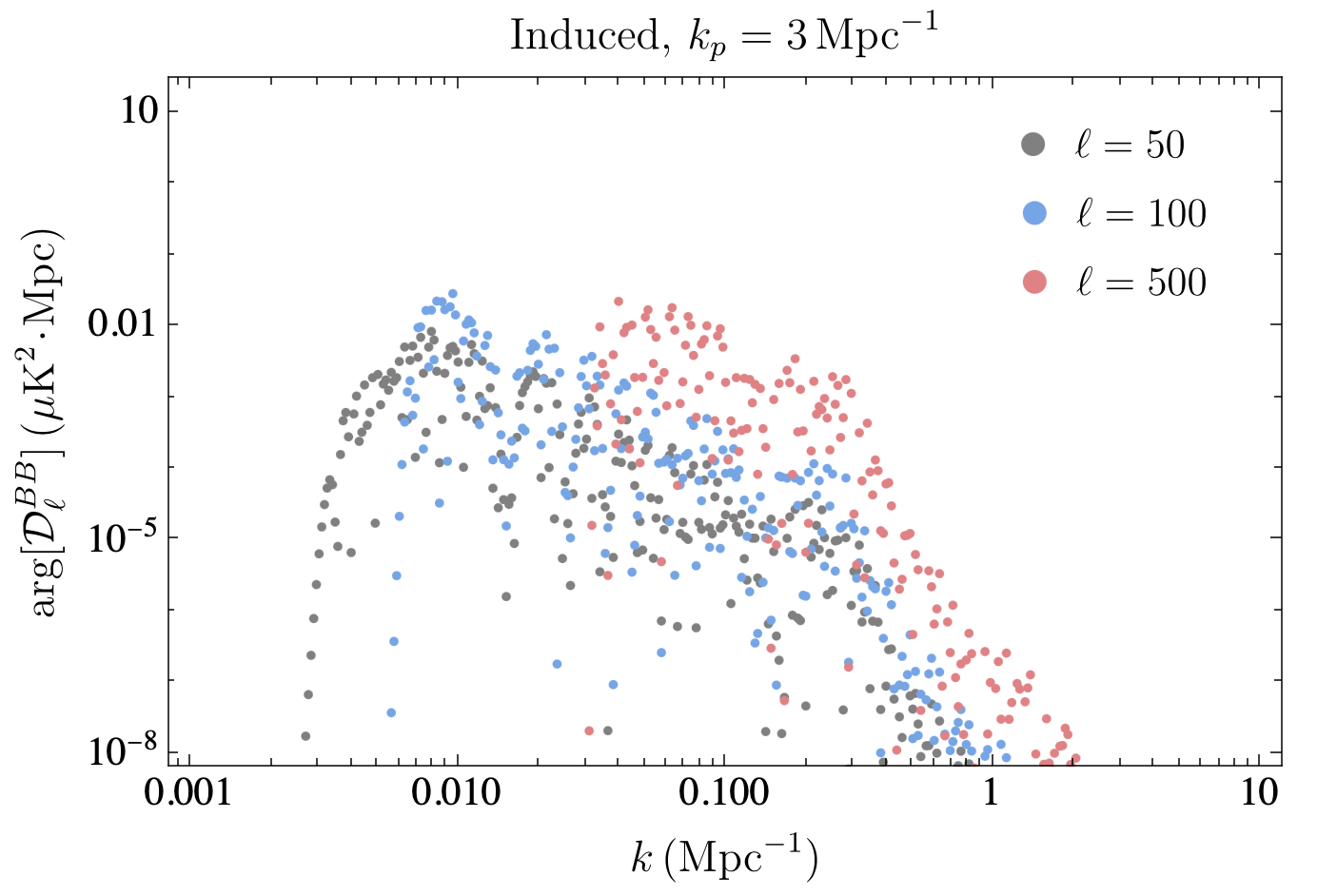}
\hspace{3mm}
\includegraphics[width=0.6\textwidth]{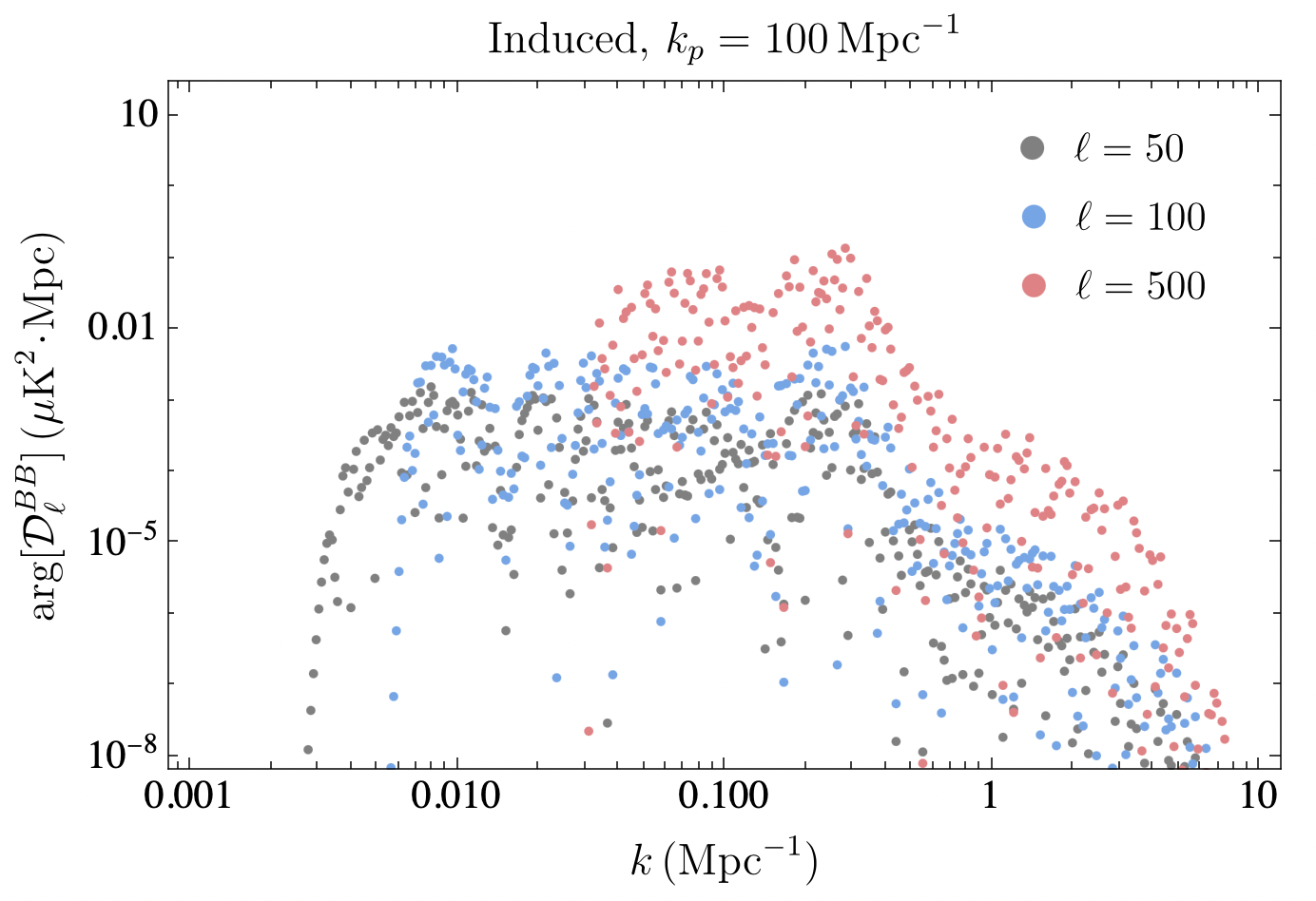}
\caption{Argument of $\mathcal{D}_\ell^{BB}$ given by Eq.~(\ref{eq:argDlBB}) as a function of $k$ for multipoles $\ell = 50$ (gray), $\ell = 100$ (light blue), and $\ell = 500$ (light red). The top panel shows this quantity for an inflationary tensor power spectrum with tensor-to-scalar ratio $r = 10^{-3}$, the middle panel shows this for an induced tensor power spectrum with $k_p = 3 \, \text{Mpc}^{-1}$ and amplitude chosen to match the inflationary signal-to-noise ratio, and the bottom panel shows the same for a peak at $k_p = 100 \, \text{Mpc}^{-1}$. The behavior in these plots illuminates the origin of the spectral shapes in Fig.~\ref{fig:DlBB}.}
\label{fig:spectralshape}
\end{figure}
The shapes of the spectra can be understood with the aid Fig.~\ref{fig:spectralshape}, wherein we plot the argument of $\mathcal{D}_\ell^{BB}$ as a function of $k$ for sample angular multipoles $\ell$. By combining Eqs.~(\ref{eq:ClBB}) and (\ref{eq:DlBB}), this argument can be written
\begin{equation}\label{eq:argDlBB}
    \text{arg}[\mathcal{D}_\ell^{BB}] = \frac{18 \ell(\ell+1) T_0^2}{k} \mathcal{P}_h(k) \mathcal{F}_\ell(k)^2 \,,
\end{equation}
which involves the product of the tensor power spectrum $\mathcal{P}_h$ and the window function $\mathcal{F}_\ell^2$. In the top panel, we plot this quantity for an inflationary power spectrum with $r = 10^{-3}$. Because the inflationary spectrum is approximately scale invariant, all of the shape dependence in this case comes from the window function, as can be seen by comparing the top panel with Fig.~\ref{fig:Flsq}. We see that the argument is maximal for $\ell \sim 100$, which is consistent with the location of the recombination peak in Fig.~\ref{fig:DlBB}.

In the middle and bottom panels, we plot this quantity for induced tensor power spectra with peaks at $k_p = 3 \, \text{Mpc}^{-1}$ and $100 \, \text{Mpc}^{-1}$, respectively, and amplitudes chosen to yield the same S/N as the inflationary signal with $r=10^{-3}$. In this case, there is competition between $\mathcal{P}_h$, which generically peaks on smaller scales (larger $k$) and $\mathcal{F}_\ell^2$, which generically peaks on larger scales (smaller $k$). The exact location of the window function's peak is $\ell$-dependent, with the peak moving to larger $k$ for larger angular multipoles (see Fig.~\ref{fig:Flsq}). Thus in this case the angular spectrum of B-mode polarization peaks at larger $\ell$ in comparison with the inflationary case. This is more dramatic for the latter benchmark.

\section{Observational Prospects}
\label{sec:prospects}


\subsection{CMB Experiments}

There are a number of ongoing and upcoming CMB experiments which aim to improve measurements of CMB B-mode polarization beyond the previous efforts of SPTpol~\cite{SPTpol:2013,SPT:2019nip}, ACTpol~\cite{ACTPol:2016}, POLARBEAR~\cite{POLARBEAR:2014}, BICEP/Keck~\cite{BICEP:2021}, and many others. Over the next decade, a combination of ground-based experiments including the BICEP Array~\cite{Hui:2018}, the Simons Array experiment~\cite{POLARBEAR:2015}, the Simons Observatory~\cite{SimonsObservatory:2018,SimonsObservatory:2025}, and CMB-Stage 4 (CMB-S4)~\cite{CMBS4,CMBS4:2020} as well as the space-based LiteBIRD~\cite{LiteBIRD:2022} and other missions will dramatically improve sensitivity to B-modes across a range of angular scales.  

Historically, the target of such experiments has been the inflationary background of gravitational waves, the magnitude of which is conveniently parameterized by the tensor-to-scalar ratio $r$. Currently, this quantity is constrained as $r < 0.036$ at $2\sigma$ based on measurements by the BICEP/Keck Array~\cite{BICEP:2021}. This constraint can be somewhat tightened to $r < 0.032$ at $2\sigma$ by using a combination of BICEP/Keck 2018 and \textit{Planck} PR4 data~\cite{Tristram:2021}.\footnote{See also \cite{deBelsunce:2022yll} for a re-analysis of the PR4 constraint using only low multipole ($2 \leq \ell \leq 30$) data.} These future experiments aim to constrain $r$ at a level $10^{-3}$~\cite{CMBS4:2020,LiteBIRD:2022} while future proposals have the even more ambitious goal of $10^{-4}$~\cite{NASAPICO:2019}.

However, as first realized in Ref.~\cite{Greene:2024}, there may exist other primordial sources of B-mode polarization that can complicate an inflationary interpretation for future measurements. For example, cosmological phase transitions and cosmic strings can also source tensor modes which are later reprocessed into CMB B-modes. In this work, we have focused on the contribution from scalar-induced tensor perturbations. As can be seen in Fig.~\ref{fig:DlBB}, the contribution from this source has a distinct shape from the scale invariant inflationary contribution, and so can in principle be distinguished provided appropriate de-lensing and measurements across a range of angular scales. 

\begin{figure}[t!]
\centering
\includegraphics[width=0.9\textwidth]{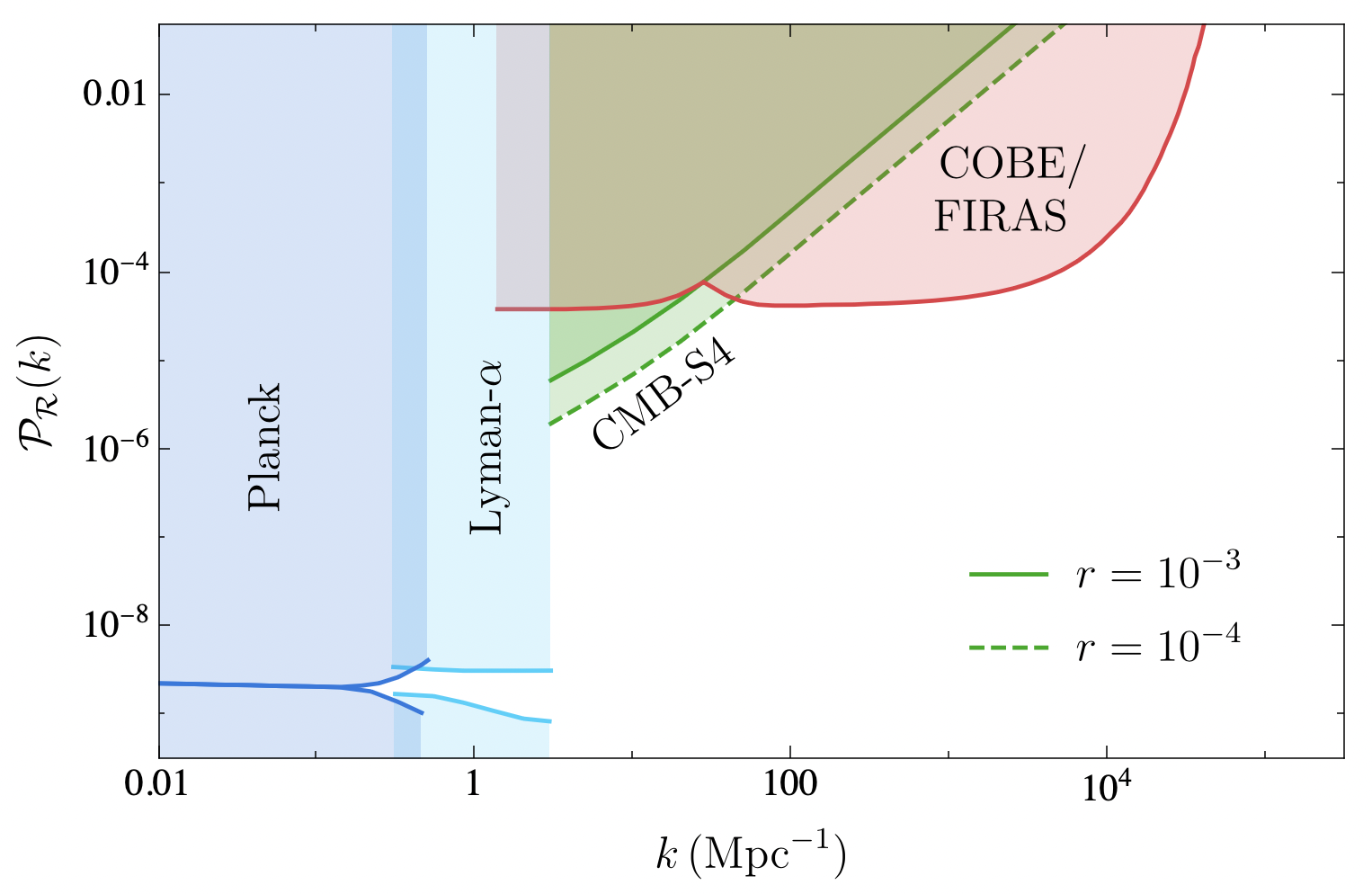}
\caption{Region of the primordial scalar power spectrum $\mathcal{P}_\mathcal{R}$ corresponding to scalar perturbation amplitudes that generate induced tensor perturbations capable of producing B-mode polarization signals detectable by next-generation experiments with target sensitivities of $r = 10^{-3}$ (solid green and above) and $r = 10^{-4}$ (dashed green and above). Solid and dashed contours are obtained by modeling $\mathcal{P}_\mathcal{R}$ by Eq.~(\ref{eq:PRbox}) with width $\Delta = 10^{-2}$ and amplitude chosen to yield the same signal-to-noise ratio as the inflationary signal with either $r = 10^{-3}$ or $10^{-4}$, presuming noise spectra corresponding to CMB-S4 forecasts~\cite{CMBS4:2020}. This region of future sensitivity extends beyond current constraints on $\mathcal{P}_\mathcal{R}$ coming from CMB spectral distortions (red)~\cite{COBEFIRAS1,COBEFIRAS2,Chluba:2012we} for $k \sim \mathcal{O}(1-10) \, \text{Mpc}^{-1}$. We show also constraints coming from primary CMB anisotropies measured by Planck (dark blue)~\cite{Planck2018,PlanckInf:2018} and Lyman-$\alpha$ forest (light blue)~\cite{LymanAlphaPk}.}
\label{fig:Pkconstraints}
\end{figure}
It turns out that the scalar amplitudes required to produce B-mode signals competitive with inflationary spectra saturating our current bounds on $r < 0.032$ are already constrained by either primary CMB anisotropies~\cite{Planck2018,PlanckInf:2018}, Lyman-$\alpha$ forest~\cite{LymanAlphaPk}, or CMB spectral distortions~\cite{COBEFIRAS1,COBEFIRAS2}. While current CMB experiments are limited in their sensitivity to small-scale B-modes, upcoming experiments like CMB-S4, the Simons Observatory, and LiteBIRD are expected to achieve improved sensitivity and resolution, making the detection of such a signal possible. 

In addition, there are a number of proposals for future experiments which would be particularly well-suited for probing this scenario. These include PICO, a proposed next-generation space-based mission aiming to detect or rule out primordial gravitational waves at a level $r \sim 10^{-4}$~\cite{NASAPICO:2019}; PIXIE, another proposed space mission intended to detect CMB B-modes as well as spectral distortions with dramatically improved precision~\cite{Kogut:2024}; and CMB-HD, a ground-based experiment whose goal is to study the CMB with higher angular resolution and wider sky coverage, focusing on small scale structures and lensing~\cite{CMB-HD:2022}. This latter proposal is especially ideal for our multipole range of interest since among its science goals is to remove 90\% of lensing B-modes over a substantial sky fraction, enabling other CMB experiments with small aperture telescopes like CMB-S4 to achieve their goals. The potential existence of our signal serves as an additional motivation for such experiments.

A detection or non-observation of scalar induced B-modes in these future experiments would allow us to infer information about the primordial scalar power spectrum $\mathcal{P}_\mathcal{R}$. In Fig.~\ref{fig:Pkconstraints}, we depict in green the region of the scalar power spectrum to which future B-mode experiments will be sensitive, presuming a target sensitivity of $r = 10^{-3}$ (solid green) or $r = 10^{-4}$ (dashed green). Concretely, these contours are obtained by modeling $\mathcal{P}_\mathcal{R}$ by the boxcar function of Eq.~(\ref{eq:PRbox}) with small width $\Delta = 10^{-2}$ and amplitude chosen such that the signal-to-noise matches that of the inflationary signal with $r = 10^{-3}$ or $10^{-4}$. We again use noise curves corresponding to future CMB-S4 forecasts~\cite{CMBS4:2020}. This future sensitivity region is plotted alongside current constraints on $\mathcal{P}_\mathcal{R}$ coming from primary CMB anisotropies measured by Planck~\cite{Planck2018,PlanckInf:2018} (dark blue), Lyman-$\alpha$ forest~\cite{LymanAlphaPk} (light blue), and CMB $\mu$- and $y$-type spectral distortions measured by COBE/FIRAS~\cite{COBEFIRAS1,COBEFIRAS2} (red). The non-observation of CMB B-modes beyond the lensing contribution in these future experiments would allow us to constrain the amplitude of $\mathcal{P}_\mathcal{R}$ beyond limits from spectral distortions for wavenumbers in the $\mathcal{O}(1-10) \, \text{Mpc}^{-1}$ range. 

Of course, ascertaining whether a measured B-mode signal originates from scalar-induced GWs will be a challenge in itself. If a signal beyond the lensing contribution is detected at a single scale, multi-messenger methods will be necessary to determine its origin. The stochastic GW background is one potential cross-check; CMB spectral distortions are another (see e.g.~\cite{Cyr:2023pgw}). Alternatively, if one can perform the measurements across a range of angular scales, then the origin can be deduced on the basis of the characteristic spectral shape.

\subsection{Stochastic GW Background}

The same scalar-induced tensor perturbations which imprint B-mode polarization on the CMB also contribute to the stochastic GW background. Thus, one can also search for these induced tensor modes directly using dedicated GW observatories.\footnote{See also \cite{Namikawa:2019tax,Ng:2021waj} for related work constraining the stochastic GW background based on CMB data.} While such searches have been proposed for existing detectors like LIGO-Virgo-KAGRA \cite{KAGRA:2013rdx} and pulsar timing arrays (PTA) \cite{NANOGrav:2023gor, EPTA:2023fyk, Reardon:2023gzh, Xu:2023wog}, they typically target GWs generated before Big Bang nucleosynthesis, since the corresponding frequencies fall within the sensitivity bands of these experiments. Similarly, GW searches using astrometric measurements~\cite{Moore:2017ity, Garcia-Bellido:2021zgu} (e.g. from Gaia~\cite{Gaia:2018ydn} and THEIA~\cite{Theia:2017xtk} surveys) tend to be sensitive to frequencies around the PTA band, $f \sim \mathcal{O}(1-100) \, \text{nanoHz}$. In contrast, induced GWs sourced by curvature perturbations at $\mathcal{O}(1-10)\, \mathrm{Mpc}^{-1}$ scales, corresponding roughly to the time of recombination, peak at significantly lower frequencies of $f \sim \mathcal{O}(10-100)~{\rm femtoHz}$. 

\begin{figure}[h!]
\centering
\includegraphics[width=0.75\textwidth]{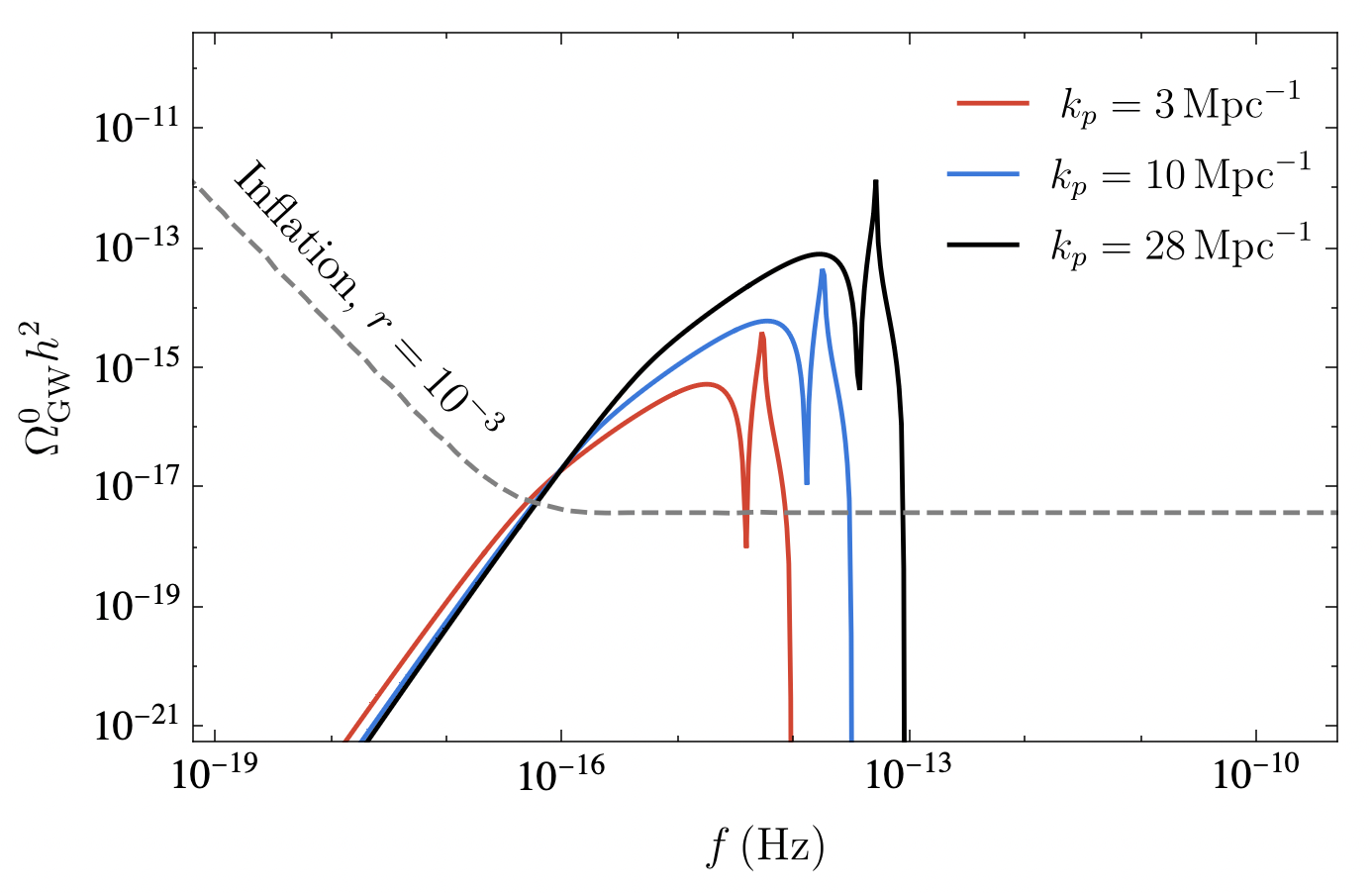}
\caption{Energy density in GWs today, parameterized by the spectral density parameter $\Omega_{\rm GW}^0 h^2$, for three benchmark points along the solid green $r=10^{-3}$ curve of Fig.~\ref{fig:Pkconstraints}. Note that the peak positions lie far below the frequencies accessible to current and future GW observatories. Note also that the low-frequency tails intersect the inflationary prediction (gray dashed) around $f \sim 10^{-17} \, \text{Hz}$, which corresponds roughly to the $\ell \sim 100$ recombination peak, indicating that the B-mode amplitudes are comparable at this scale.}
\label{fig:GWspectrum}
\end{figure}

In Fig.~\ref{fig:GWspectrum}, we show the present-day energy density in GWs, as parameterized by the spectral density parameter $\Omega^0_{\rm GW} h^2$ defined in Eq.~(\ref{eq:OmegaGWtoday}), for three benchmark points from the $r = 10^{-3}$ contour of Fig.~\ref{fig:Pkconstraints}. These benchmarks feature peak locations at $3$, $10$, and $28~\text{Mpc}^{-1}$ and amplitudes chosen such that the signal-to-noise ratio matches the $r = 10^{-3}$ inflationary signal. The induced GW spectrum, like the B-mode polarization angular spectrum, peaks at small scales corresponding to the horizon re-entry time of the scalar perturbations. Note also that the low-frequency tails of all three curves intersect the inflationary curve (gray dashed) around $f\sim 10^{-17}~{\rm Hz}$, or equivalently $k \sim 0.01~{\rm Mpc}^{-1}$ and $\ell \sim 100$, indicating that the B-mode polarization signals are all comparable at this scale.

While all three benchmarks are expected to be accessible in the next generation of CMB polarization experiments, they lie far outside the frequency range probed by current and future GW observatories. This underscores a key advantage of our proposal: by reprocessing tensor perturbations into B-mode polarization, it becomes possible to access scales which would otherwise be out of reach using traditional GW detection methods. 


%
%

\section{Conclusion}\label{sec:conclusions}

In this work, we have presented a novel probe of small-scale curvature perturbations based on scalar-induced tensor perturbations and their imprint on the polarization spectrum of the CMB. Presuming a localized enhancement of the scalar power spectrum, we have calculated the induced tensor spectrum at horizon crossing, evolved it using appropriate transfer functions, and computed the contribution to the angular spectrum of B-mode polarization. We found that the resulting spectral shape was generically distinct from that predicted from inflationary tensor perturbations, with more power on smaller scales and a non-existent reionization peak. For various scalar peak positions and amplitudes, we computed the signal-to-noise ratio and compared with the sensitivity of current and future CMB experiments, mapping the projected sensitivities back onto the scalar power spectrum. 

While current sensitivities to small scale B-modes are limited, upcoming CMB polarization surveys like CMB-S4~\cite{CMBS4,CMBS4:2020}, the Simons Observatory~\cite{SimonsObservatory:2018,SimonsObservatory:2025}, and LiteBIRD~\cite{LiteBIRD:2022} are expected to improve angular resolution such that a detection may be possible. The prospect of a detection also motivates dedicated CMB B-mode searches at larger angular multipoles $\ell \gtrsim 100$, strengthening the scientific case for future proposals like PICO~\cite{NASAPICO:2019}, PIXIE~\cite{PIXIE,Kogut:2024}, and CMB-HD~\cite{CMB-HD:2022}. Regardless of whether or not B-modes beyond the lensing signal are ultimately observed, these experiments will provide powerful information about curvature perturbations on small and intermediate scales. In particular, a non-detection would place the strongest existing limits on the primordial power spectrum for scales $k \sim \mathcal{O}(1-10)\, \mathrm{Mpc}^{-1}$. These new constraints would be both robust and generic. Unlike other methods attempting to probe the small-scale primordial power spectrum, which suffer from modeling uncertainties in small-scale structure and astrophysics, tensor perturbations offer a remarkably clean, statistically distinct signal. 

There are a number of future directions which would be interesting to pursue. Our result has built on the research program initiated in Ref.~\cite{Greene:2024}, which aimed to identify sources of primordial tensor modes beyond those from inflation and study their observable consequences in CMB polarization. While this earlier study examined a particular class of cosmological phase transitions, and here we have investigated scalar-induced GWs, many additional sources remain to be explored --- including cosmic string networks and other classes of cosmological phase transitions. Within the induced GW scenario itself, there remain several avenues for further study. We have limited ourselves to the scenario of curvature perturbations with a Gaussian probability distribution, however it would be worthwhile to also investigate non-Gaussian induced GWs~\cite{Cai:2018,Perna:2024}, since in this scenario one can potentially achieve even larger tensor power with smaller enhancements in the scalar power spectrum. The non-observation of B-modes --- even at current sensitivity levels --- could allow one to constrain the non-linearity parameters $f_{\rm NL}$ and $g_{\rm NL}$ at smaller scales than existing CMB constraints.

\acknowledgments KS would like to thank the Aspen Center for Physics, which is supported by National Science Foundation grant PHY-2210452, for hospitality during the course of this work. KS and TX also wish to acknowledge the Center for Theoretical Underground Physics and Related Areas (CETUP*) and the Institute for Underground Science at SURF for hospitality and for providing a stimulating environment. AI is supported by NSF Grant PHY-2310429, Simons Investigator Award No.~824870, DOE HEP QuantISED award \#100495, the Gordon and Betty Moore Foundation Grant GBMF7946, and the U.S.~Department of Energy (DOE), Office of Science, National Quantum Information Science Research Centers, Superconducting Quantum Materials and Systems Center (SQMS) under contract No.~DEAC02-07CH11359.  The research activities of KS and TX are supported in part by the U.S. National Science Foundation under Award No. PHY-2412671.

\bibliographystyle{JHEP.bst}
\bibliography{biblio}

\end{document}